\documentclass{aa}
\usepackage{graphicx}   
\newdimen\captwidth   
\newdimen\figwidth   
\def\eg{{\it e.g. }}
\def\etal{{et al. }}
\def\ie{{\it i.e. }}
\def\micron{\hbox{$\,\mu {\rm m}\,$}}
\def\microJy{\hbox{$\,\mu {\rm Jy}\,$}}

\def\zu{\rm\,}     
\def\te{\rm\ }     

\begin{document}
   \thesaurus{02         
     (03.13.2;  
      11.05.2;  
      11.16.1;  
      11.19.7;  
      12.03.3;  
      13.09.1)  
            }
%
 
   \title{A classical approach to faint extragalactic source extraction
from ISOCAM deep surveys\thanks{ Based on observations with ISO, an
  ESA project with instruments funded by ESA Member States (especially
  the PI countries: France, Germany, the Netherlands and the United
  Kingdom) and with participation of ISAS and NASA}
}
 
\subtitle{Application to the Hubble Deep Field}
 
   \author{F.--X. D\'esert \inst{4, 1}  \and J.--L. Puget \inst{1} \and 
D. L. Clements \inst{1} \and  
M. P\'erault \inst{2, 1} \and A. Abergel \inst{1} \and J.--P.
Bernard \inst{1} \and C. J. Cesarsky \inst{3}}
 
   \offprints{F.--X. D\'esert$^4$}
 
   \institute{ Institut d'Astrophysique Spatiale, B\^at.
121, Universit\'e Paris XI, F--91405 Orsay Cedex France, name@ias.fr
\and
               Groupe de Radioastronomie millim\'etrique, ENS,
24 rue Lhomond, F--75005 Paris France, perault@ensapa.ens.fr
\and
            Service d'Astrophysique/DAPNIA/DSM, CEA-Saclay, 
l'Orme des Merisiers, F-91191 Gif-sur-Yvette  Cedex France, delbaz@cea.fr
\and
    Laboratoire d'Astrophysique, Observatoire de Grenoble BP 53, 
    414 rue de la piscine, F--38041 Grenoble Cedex 9  France\\
    e-mail: Francois-Xavier.Desert@obs.ujf-grenoble.fr
             }
 
\date{Received Jan. 1998/ Accepted 17 Aug. 1998}
\titlerunning{Source extraction from ISOCAM deep surveys} 
   \maketitle

   \begin{abstract} 
     
     We have developed a general data reduction technique for ISOCAM
     data coming from various deep surveys of faint galaxies. In order
     to reach the fundamental limits of the camera (due to the
     background photon noise and the readout noise), we have devised
     several steps in the reduction processes that transform the raw
     data to a sky map which is then used for point source and
     sligthly extended source extraction. The main difficulties with
     ISOCAM data are the long-term glitches and transient effects
     which can lead to false source detections or large photometric
     inaccuracies. In many instances, redundancy is the only way
     towards clear source count statistics. A sky pixel must have been
     ``seen'' by many different CAM pixels. Our method is based on
     least-squares fits to temporal data streams in order to remove
     the various instrumental effects. Projection onto the sky of the
     result of a ``triple--beam method'' (ON -(OFF1 + OFF2)/2)
     obtained from the signal of a given pixel during three
     consecutive raster positions leads to the removal of the low
     frequency noise.  This is the classical approach when dealing
     with faint sources on top of a high background. We show
     illustrative examples of our present scheme by using data taken
     from the publicly available Hubble Deep Field ISOCAM survey in
     order to demonstrate its characteristics.
     
     More than thirty sources down to the 60 (resp. 100) \microJy
     \allowbreak level are clearly detected above 4 $\sigma$ at
     wavelengths of 7 (resp. 15) \micron, the vast majority at 15
     \micron.  A large fraction of these sources can be identified
     with visible objects of median magnitude 22 and K--band magnitude
     of 17.5 and redshifts between 0.5 and 1 (when available).  A few
     very red objects could be at larger redshifts.


   \end{abstract}
 
\section{Introduction}

After the first ISOCAM results on relatively strong sources (ISO A\&A
1996, 315, special issue) it has become clear that the Camera on ISO
could achieve more sensitive observations by integrating longer,
particularly on ``empty'' high galactic latitude fields. Already
sources with fluxes larger than the mJy level can routinely be
observed which are 1000 times fainter than previous IRAS 12$\micron$
surveys and with a resolution which is 10 times higher. Reduction of
the data requires that effort has to be spent on dealing with CAM
peculiarities which show up at these levels. The now ``old''
mid-infrared detectors on ISO, which were designed ten years ago,
suffer from long--term memory effects that have to be dealt with,
before the noise can be integrated down. We will show that this is
possible because the camera is ultimately highly linear when
stabilised, and the so--called transient effects are highly
reproducible under the same conditions, and because the zodiacal
background, which dominates the high galactic latitude sky in the long
wavelength (LW) channel, appears to be smooth on the 10 arcsecond
scale. As many important projects with ISOCAM (several hundred hours
from both guaranteed and open time proposals) aim at looking for faint
extragalactic objects, it is worth developing and demonstrating
reliable, independent data reduction techniques. Our approach is
complementary to the method presented by Starck \etal (1998) and
Aussel \etal (1997, 1998).  We will here present how the raw data are
processed to remove the main instrumental effects
(Sect.~\ref{se:proc}), and how the triple beam--switch reduced data
are projected onto the sky and sources are extracted from the
resulting sky map (Sect.~\ref{se:proj}). At each step, for
illustrative purposes, we take examples from the HDF public domain
ISOCAM data (Rowan--Robinson \etal 1997).  All of our algorithms were
developed on Dec-Alpha stations using specific routines written in the
IDL language. We then apply the whole ISOCAM data reduction method to
the HDF specific case and give a list of the faintest mid-infrared
sources that have been observed to date (Sect.~\ref{se:hdf}).
Finally we briefly discuss the advantages and drawbacks of this method
and the various ways it could be improved.

\section{The ISOCAM data processing}\label{se:proc}

\subsection{Reading the raw data}
\label{ss:rd}

\begin{figure*}[tb]    
  \includegraphics[width=\textwidth]
    {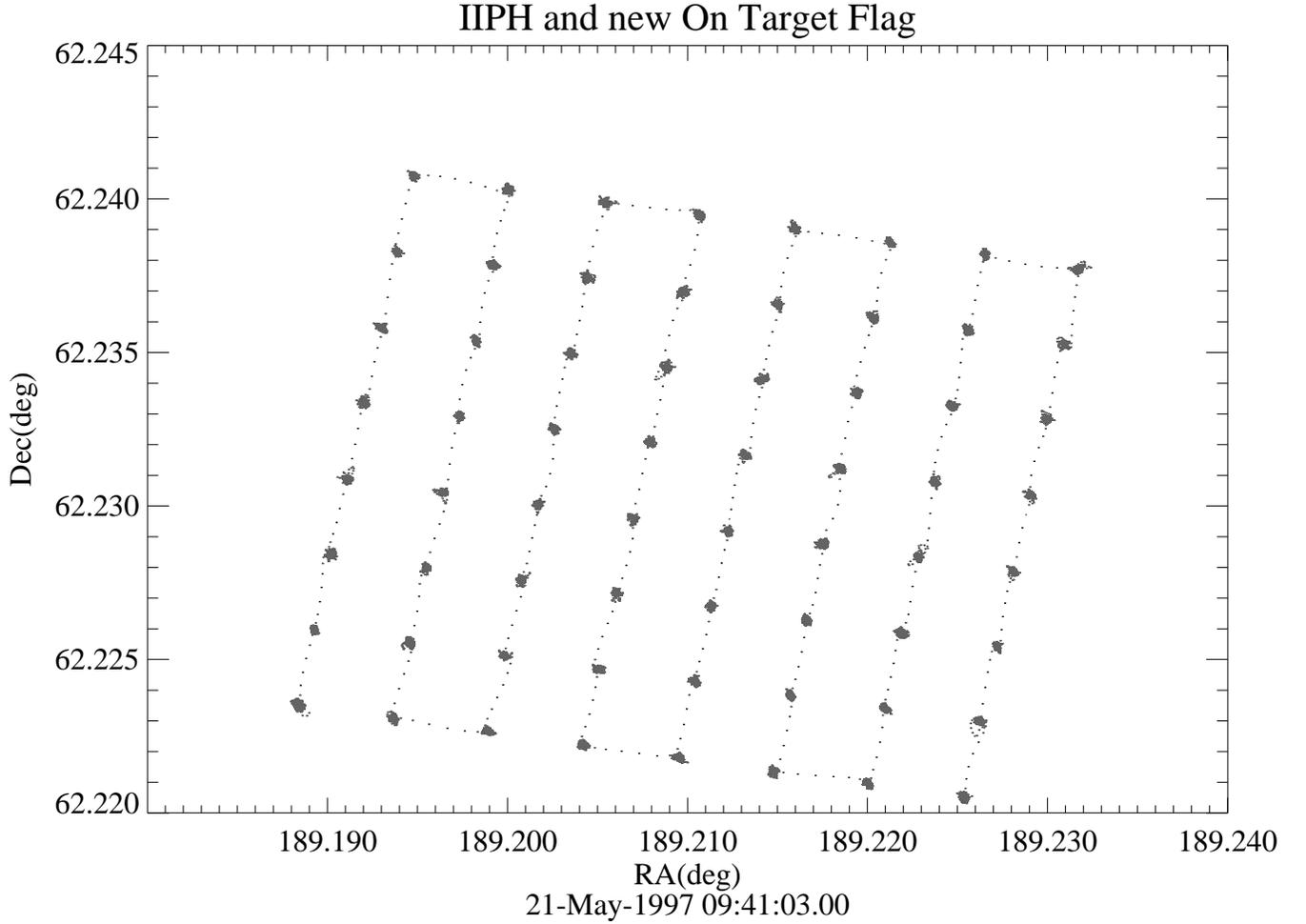}
  \caption{ISO pointing history taken from
 IIPH22701702. Consecutive points are separated in time by 
half a second. The positions in RA and Dec (2000) are smoothed 
with a 2 second temporal kernel. One must mask out readouts which were taken
when during some part of them, ISO was moving.}
  \label{fi:iiph}
\end{figure*}

Raw telemetry from ISO is converted into different formats and delivered by ESA
to the community.  We chose to start from SPD FITS files. This means that the
pair of reset and end-of-integration images are already subtracted. The CISP
(almost raw) data are read into a dataset which contains, along with the data
cube (of size 32 by 32 by the total number of readouts), a header describing
the context of the observations, and various trend parameter arrays (CAM wheel
positions, various ISO time definitions) which are or are not synchronised with
the images. An important trend parameter set is the pointing information from
the ISO satellite which is read from the appropriate IIPH (ISO Instrument
Pointing History) FITS file and appended to the raw dataset. The equatorial
right ascension (RA) and declination (Dec) coordinates (in the J2000 system)
pointed to by the centre of the camera suffer from some noise (\emph{not}
corresponding to a real ISO jitter). These are smoothed with a running kernel
with a FWHM of 2 arcseconds which give a relative pointing accuracy better than
one arcsecond  (Fig.~\ref{fi:iiph}). The pointing data are synchronised to the
dataset cube of images by using the UTK time that is common to CISP and
IIPH files. Data are usually taken in a raster mode where a regular grid of
pointings on the sky is done successively. Care should be taken not to use the
on-target flag delivered in the CISP file. This flag is set to one when the
acquired pointing is within 10 arcsecond of the required raster pointing. This
flag is obviously too loose if one uses the 6 or, even worse, the 3 arcsecond
lens. A 2 arcsecond radius of pointing tolerance is used here for our own
criterion.

At this stage the data in the ISOCAM internal units of ADU
(Analog--to--Digital Units) are converted to
ADUG units by simply dividing by the gain (equal to $2^{\tt
F2\_ADC\_GAIN}$). A library dark image is then subtracted, the accuracy of
which is not relevant in the following. In the next three sub--sections
we consider each pixel as individual detectors and analyse their
temporal behaviour without regard to their neighbours. A mask cube of
the same size as the data cube is used in parallel to flag the data
which are affected by various damaging effects.
 
\subsection{Removing particle hits}
\label{ss:slg}

\subsubsection{Fast glitch correction}
\label{sss:fgc}

\begin{figure*}[htbp]    
  \includegraphics[width=\textwidth]
  {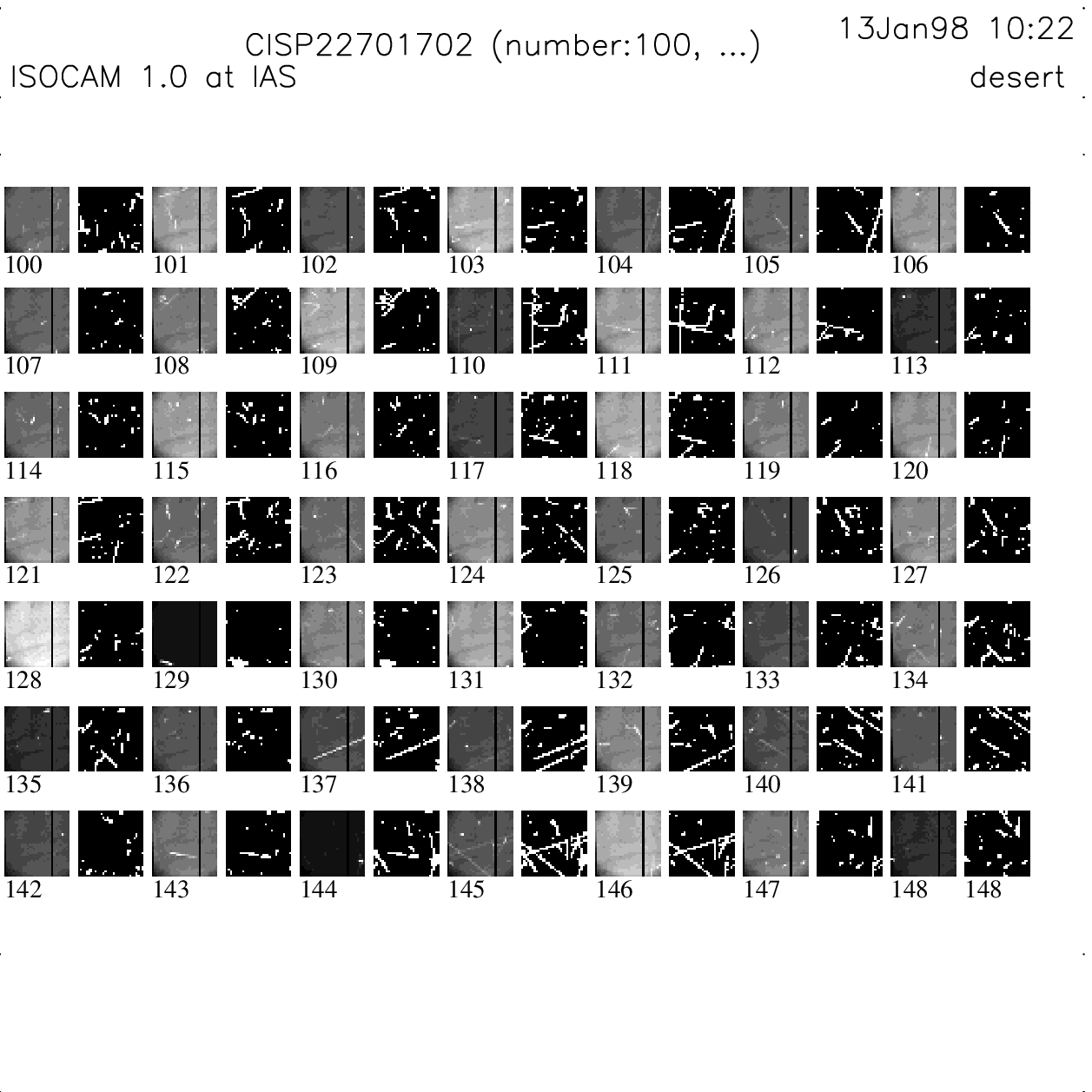}
  \caption{A sample of elementary successive images with the
    ISO 32 by 32 pixel LW camera taken from the CISP22701702 data set,
    number 100 to 148 (starting at 0). After each image, a mask
    deduced from the short deglitching algorithm shows the numerous
    impacts of cosmic rays.  (Configuration of the camera is LW3
    filter, LGe6 lens, 36 camtu (=5 seconds) of elementary integration
    time.)}
  \label{fi:CR1}
\end{figure*}
At first glance (Fig.~\ref{fi:CR1}), any readout obtained with ISOCAM
not containing strong sources (fluxes larger typically than about 10
mJy in the broad filters) clearly shows the flat field response by the
pixels to the zodiacal background. Ultimately, it is the accuracy to
which one knows this flat--field that allows identification of faint
sources (see Sect.~\ref{ss:LTD}).

\begin{figure*}[tbhp]    
  \includegraphics[width=\textwidth]
  {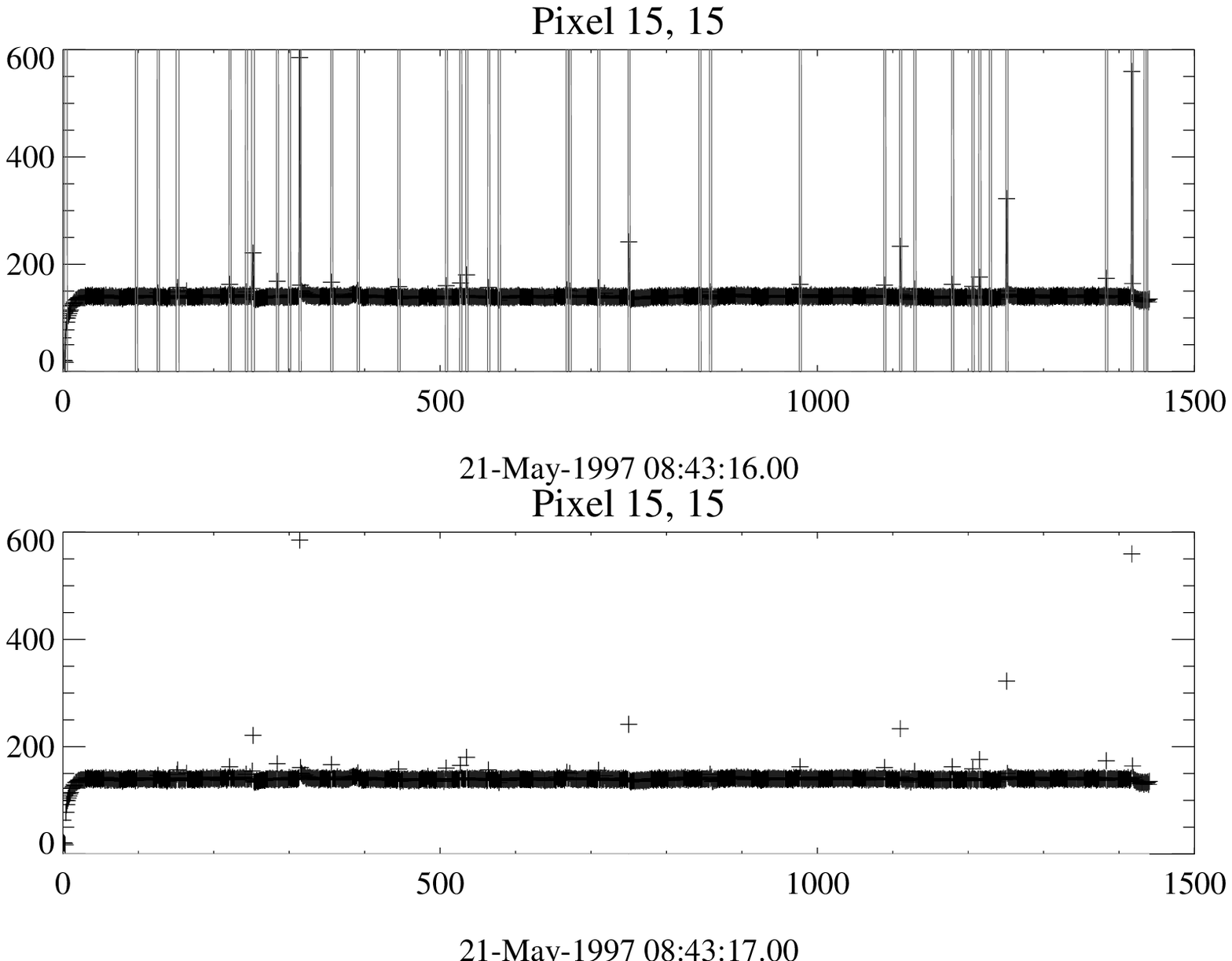}
  \caption{Temporal evolution of the raw signal (in ADUG \ie
    Analog--to--Digital Units divided by the gain) from camera pixel
    15, 15 (starting at 0, 0) taken in CISP22701702 (abscissa is in
    readout number starting at 0 for the first readout of the
    dataset).  {\bf Upper panel} -- A mask of readouts affected by
    fast glitches as deduced by the algorithm described in the text
    (dotted vertical lines).  {\bf Lower panel} -- the raw data}
  \label{fi:CR2}
\end{figure*}

Detectable signals from some pixels which are often aligned in strings
result from hits by cosmic ray particles. Most of the affected pixels
recover at the next or second next readout. These are mainly due to
primary and secondary electron energy deposition onto the array.

The affected pixels are found by an algorithm working on the temporal
behaviour of each pixel. Readouts of one pixel which deviate from the
running (14 readouts) median by more than a threshold value and tend
to recover to the normal level by at least 10 percent of the maximum
step after one or two readouts are simply masked for two readouts and
excluded from further signal extraction.  The threshold value is set
by a number (typically 3) times a running window (14 readouts)
standard deviation of the pixel signal, where the most deviant values
have been excluded from the rms computation.  This algorithm was
tested in various cases, and in particular against false glitch
detection around the typical signal of a moderately strong source.
Figure~\ref{fi:CR2} shows an example of the temporal behaviour of one
CAM pixel as a function of readout number.

The number of glitches, or more precisely, the number of readouts
which are affected by the masking process, is typically 9 per second
over the entire array of 992 alive pixels: \eg during an integration
time of $t_{int}$ = 36 camtu = 5 seconds, at any time 45 pixels cannot
be used for measurement.  This number varies by at least 15 percent
apparently depending on the satellite orbiting position. Accumulated
glitches cause a noticeable increase in the noise or an unreliable
measurement. We thus decided to mask an isolated ``good'' readout
value if it is between two successive glitches. The energy deposition
has a continuum distribution that goes down to the intrisic noise
level of the camera. For the method presented here, any undetected
faint glitch contributes to increase the statistical noise and to
modify slightly the signal.

\subsubsection{Slow glitch correction}
\label{sss:sgc}

\begin{figure*}[tbhp]    
  \includegraphics[width=\textwidth]
  {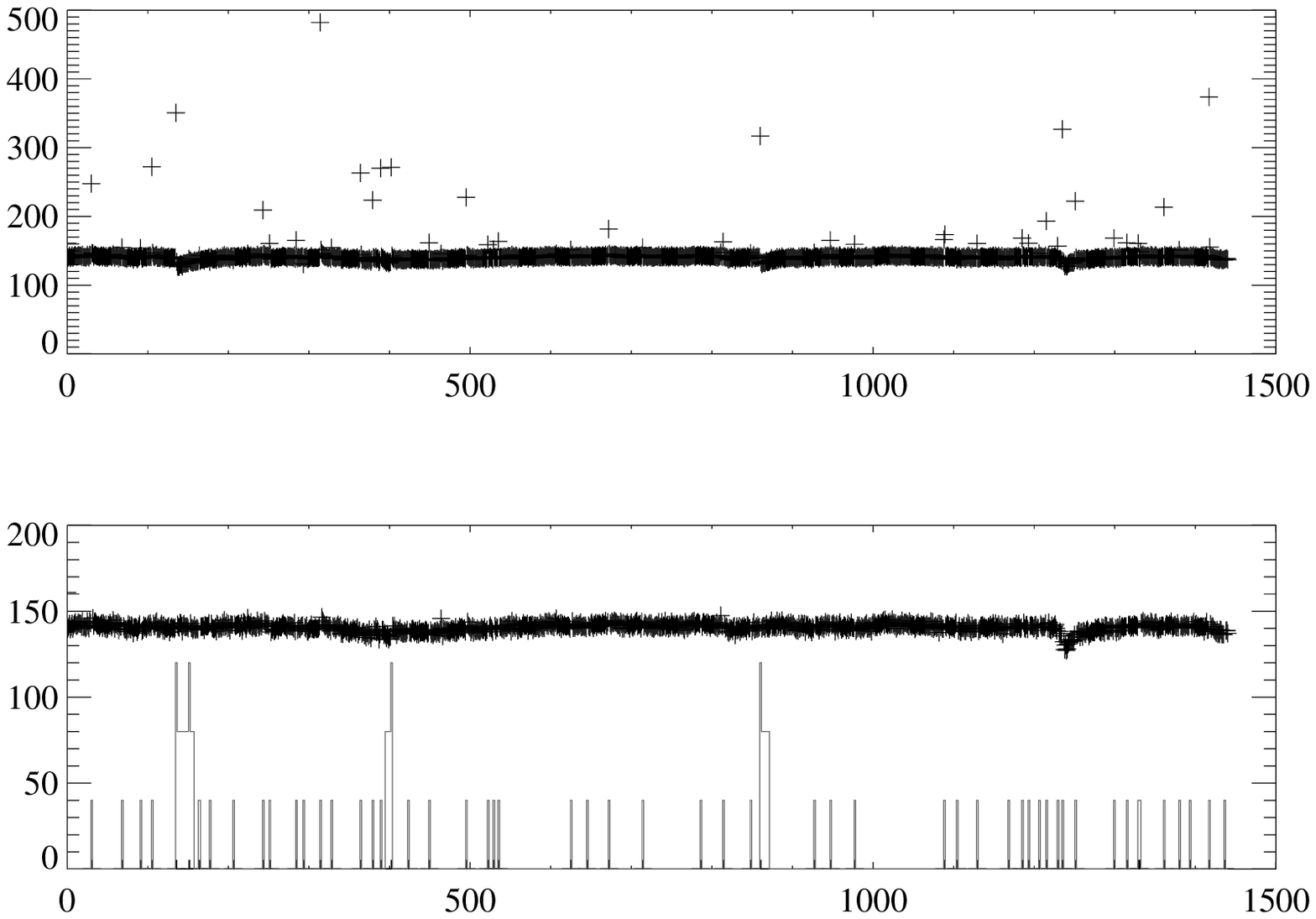}
  \caption{Temporal evolution of the raw signal (in ADUG) from 
camera pixel 16, 15 (starting at 0, 0) 
taken in CISP22701702 (abscissa is in readout number starting
at 0 for the first readout of the dataset).
{\bf a} -- Raw signal (except for the removal of the first camera
overall transient which can be seen in the previous figure)
{\bf b} -- 
The signal after the removal of slow glitches and the adapted mask
(40 is for a fast glitch, see Sect.~\ref{sss:fgc}, 
and 80 for a slow glitch,see Sect.~\ref{sss:fgc}): a non
vanishing mask corresponds to a value which will not be used for further
processing, masked values are not shown. The dip near readout 1250 is
not corrected at this stage.}
  \label{fi:slg}
\end{figure*}

These are the most difficult to deal with. They are the main
limitation in detecting weak point-like or extended sources and are
thought to be due to ion impacts. These impacts affect the response of
the hit pixels with a long memory tail.  Slow glitches are of three
types: 1) Positive decay of short time constant (5-20 seconds), 2)
Negative decay of long time constant (up to 300 seconds) 3) A
combination. It is possible that positive decay glitches are due to an
ionisation or energy deposition which does not saturate the detector.
Negative ones may have started with saturation (which is not apparent
in the readout value) and hence cause memory effects, an upward
transient of a type not very different from the usual CAM transient
starting from a low flux.

Figure~\ref{fi:slg}a shows the effect of slow glitches.  In
Fig.~\ref{fi:slg}b the corrected signal is shown along with the used
mask.  To detect slow glitches we correlate a running template of a
typical glitch along the temporal signal $D_t$ of a given pixel. A
maximum in the correlation indicates a potential glitch at say $t_0$.
This is then analysed with a least-square method in order to find the
best decay time constant: one minimises the $L_a$ quantity defined by

\begin{eqnarray}
L_a^2 & = & \Sigma_t w_t (D_{t}-M_{t})^2 {\,\rm with\,} \\
M_{t}& = & A+ Bt+ H(t-t_0) G 
\exp{\big(-{(t-t_0)\over\tau}\big)},
\end{eqnarray}
where $D_t$ is the raw temporal data signal of a pixel, $w_t$ the
weight (0 or 1, defined by the previous masking of fast glitches),
$M_t$ the glitch model to fit, and $t_0$ is given.  Hence, the
variables $A$, $B$, $G$, and $\tau$ are found ($H$ is the Heaviside
jump function).

If the fit is satisfactory, we subtract the exponential part of the
fit up to the end of the temporal signal. To assess whether the fit is
satisfactory or not, we also calculate the least-square $L_b$
corresponding to a linear baseline without the exponential part. We
found that, a potential glitch can be taken as valid if $L_a \le
L_b/2$.  Most of the time, the glitch beginning at time $t_0$
corresponds to a previously detected fast glitch.

Type 1 glitches can be removed with the previous method but,
sometimes, the same method also removes real source signals (for
example, if there is a downward transient when a pixel leaves a
source). We thus leave Type 1 glitches to be removed later in the
method.  Type 3 glitches are not dealt with at the moment. The running
template to detect a glitch and find its starting time $t_0$ is a
simple exponential with successive time constants of 15, 30, and 60
readouts for negative glitches (Type 2 glitches). After a glitch is
found at a significant level, and the exponential tail is corrected
everywhere after $t_0$, we mask (i.e. we will not further use) the
readouts for times between $t_0$ and the time $t_1$ where the
amplitude of the exponential correction is above twice the pixel noise
per readout. An example can be found in Fig.~\ref{fi:slg}.

Typically one new slow negative glitch appears somewhere on the camera
every 1.2 second. Its intensity ($G$) varies from 5 to 20 ADUG.  The
time constant varies from 20 to 200 seconds (45 is typical).  Positive
glitches are 10 times rarer.

\subsection{Removing transient effects: a simple correction technique}
\label{ss:tr}

\begin{figure*}[htbp]   
\def\fighfrac{0.8}
\figwidth=\fighfrac\textwidth
  \includegraphics[width=\figwidth, angle=90, origin=br]
   {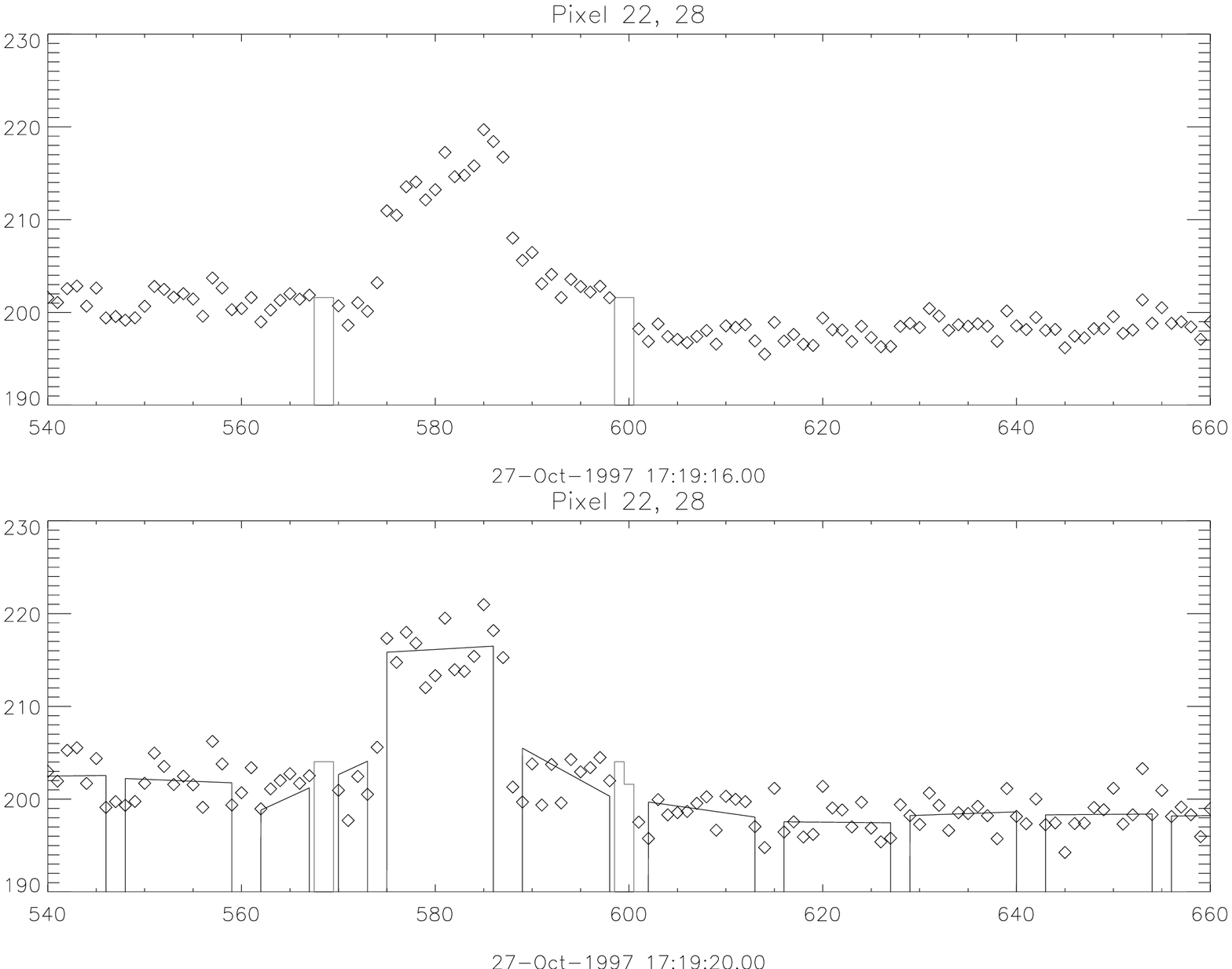}
\caption{ {\bf a} -- The transient phenomenon of ISOCAM
  is illustrated on a relatively strong source taken from ISOCAM
  guaranteed time deep surveys. The signal $D_t$ is shown in ADUG as a
  function of readout number. Eight ISO raster positions are within
  the plot. At readout 575, just after ISO moved, the pixel (22, 28)
  instantaneously responds to the illumination added by a (rather
  strong) source on top of the zodiacal background, but it also has a
  long lagged response (mask is in dotted line). After ISO moved to
  the next raster position, one can still see the memory effect of the
  source.  {\bf b} -- The recovered illumination history $I_t$ of the
  same pixel after inversion of the transient model (Eq.~\ref{eq:tr}).
  The solid line is the result of the triple--beam switch fitting
  (Eq.~\ref{eq:3pt}, only the central part of the three--leg best fit
  of each raster position is shown for simplicity). The fit is done on
  valid readouts within each raster position, as delimited by the
  vertical lines. No sources in the HDF can be seen as vividly as here
  because they are much fainter.}
\label{fi:tr} 
\end{figure*}

ISOCAM, like many other infrared detectors, suffers a lag in its
response to illumination. Fortunately, the LW detector has a
significant instantaneous response $r\simeq 0.60$ \ie a jump in
brightness is seen at once by the pixel at the level of 60\% of the
step relative to its stabilised asymptotic value.  The remainder of the
signal is obtained after a delay which is inversely proportional to
the flux. In a first approximation, Abergel \etal (1996) have modelled
this phenomenon with:

\begin{eqnarray}
D_{t} = r I_{t} + (1-r) \int^t_{-\infty} dt^\prime k_{t^\prime} I_{t^\prime}
e^{-k_{t^\prime}(t-t^\prime)}  {\,\rm and\,}\\
k_{t^\prime}= \alpha I_{t^\prime} \simeq \alpha D_{t^\prime}, 
\label{eq:tr} 
\end{eqnarray}
where the measured signal $D_{t}$ at time $t$ is a function
of the illumination
$I_{t^\prime}$ at previous times, and  
\begin{eqnarray}
\alpha= 1200\zu ADUG^{-1} readout^{-1},
\end{eqnarray}
where an ADUG is the CAM analog to digital unit normalised by the used
gain.  The approximation for $k$ in eq.~\ref{eq:tr} makes it possible
to invert the triangular temporal matrix. The inversion algorithm is
independent of the position of the satellite and makes no a priori
assumption as to the temporal evolution of the pixel intensity
history. It also preserves the volume of the data cube. An example of
a relatively strong source is shown in Fig.~\ref{fi:tr}a. The
inversion (getting $I$ from $D$) yields the result shown in
Fig.~\ref{fi:tr}b. The inversion apparently enhances the high
frequency noise of the pixel but the \emph{signal to noise ratio}
stays constant because the overall ISOCAM calibration must be updated
after the transient correction has been applied.

The correction helps in improving the calibration accuracy because it
gives the proper stabilised value that can be directly tied to a
response measured on bright stabilised stellar standards. It also
removes ghosts of sources which otherwise can still be seen after the
pixel has been pointed away from the source to its next raster
position (see Fig.~\ref{fi:tr}a).  We believe this correction works
best for faint sources, the main objective of the present study. The
correction is not yet perfect and further understanding of the camera
lag behaviour will certainly provide improvements in the final
calibration.

\subsection{Removing long term drifts: the triple beam-switch
method}\label{ss:LTD}

The data cube should now contain a signal in the unmasked areas, which
is almost constant for a given raster position and a given pixel.  At
this stage, we remove most of the slow positive glitches for a given
pixel 
\begin{enumerate}
\item by computing the standard deviation found
inside each configuration (\ie one raster position) and by taking the
median average of this deviation for all configurations.

\item by masking a readout if it deviates
by more than 4 times this typical noise from the median level of its
configuration
\end{enumerate}
The value of 4 was found by trial and error.

We are then left with a data cube where all or almost all ``bad''
pixels have been masked and therefore are not used further. As there
is still some low-frequency noise for each pixel, we do not feel ready
yet to project the total power value of each pixel on the sky but
instead we prefer comparing the values during a raster position to the
values in the two adjacent raster positions seen by the same CAM
pixel. This is the classical approach for dealing with low-frequency
noise.  It is usually adopted when the background is much stronger
than the sources: a regime which has long been the case in infrared
astronomy and which is now appearing even in optical astronomy. This
approach has been pioneered by \eg Papoular (1983).

This is done with the following least-square method which is
independently applied to each pixel (where $i$ is the readout number
proportional to the time $t_i$ which runs along the 3 current raster
positions centered at the mid time), by minimising:

\begin{eqnarray}
Ls= \sum_i w_i(D_i-(a+bt_i+up_i))^2,
\label{eq:3pt}
\end{eqnarray}
where $w_i$ is 1 for valid readouts and 0 for masked readouts or
readouts which do not belong to the three current raster positions \eg
when ISO was moving (see Sect.~\ref{ss:rd}).  $D_i$ is the pixel
signal at readout $i$.  $p_i$ is the template of a source, typically a
square pattern (0...0, 1,...,1,0...0) template, where the ones are set
for the central raster position. The best $a$, $b$, and $u$ are
therefore obtained from the minimisation of $Ls$. The method gives an
estimate of the noise on the $a_b$, $b_b$, and $u_b$ parameters by
assuming that roughly $\sigma^2= Ls_b / (\sum_i w_i - 3)$ (because 3
parameters are fitted) so that the noise per readout is $\sigma$ and
$Ls_b$ is obtained from formula \ref{eq:3pt}.  The value of $u_b$
(simplified to $u$ in the following) corresponds to the best estimate
of the average signal for each pixel and raster position. The
associated noise $\sigma_u$ and $Ls_b$ are recorded for a given pixel
at a given raster position. The uncertainty on the pixel signal
$\sigma_u$ is itself quite noisy for a given raster position so in fact we
replace it by the median of all $\sigma_u$ found during the raster for
that particular pixel (the quality of the fit $Ls_b$ is modified
accordingly). As a complementary and very effective 
glitch removal, we mask the signal of
a given raster position if its $Ls_b$ deviates by more than a factor 2
from the pixel median value across all rasters or if the number of
points used for the fit is less than a factor of 0.6 times the median
value.  No detection of a point source is made at this stage. Hence,
the data cube is reduced to a few values per raster position and per
pixel of the camera. For the first and last positions of the raster,
we use a 2-beam differencing scheme similar to the 3-beam scheme
presented except that no slope $b$ can be found. We checked, a
posteriori, that the distribution of all the values of the signal $u$
divided by their respective noise $\sigma_u$ precisely follows a
reduced Gaussian (actually we slightly overestimate the noise by up to
15 percent), except for the few pixels affected by sources. This is
strong evidence that white noise dominates the output of this
algorithm.

Note that the standard raster averaging method followed by
ON-(OFF1+OFF2)/2 differencing scheme would have worked in most
situations except that here the noise can be estimated independently,
the least square statistics can be used as an effective glitch
removal, and, in the case of several randomly placed masked values the
baseline removal is better defined.  Note also that the noise of the
triple beam-switch method is $\sqrt{3/2}$ worse than for an absolute
measurement (in case the flat field were perfectly known). But the
low-frequency noise is here largely suppressed which overbalances the
loss of sensitivity, which in principle costs an integration time
longer by 50 percent on target.

\section{Projection onto the sky and extraction of sources}\label{se:proj}

\subsection{Projection onto the sky}\label{ss:proj}

The reduced cube ($u$) can now be flat fielded (FF). Notice that at
this stage the FF only applies to values which are close to zero and
that the precision for this FF is not critical. Actually, the method
outlined above provides a natural FF for the data, namely the median
of the fitted background (as given by the term $a_b+ b_b t_{ic}$ applied to
the central time of each raster) along all raster positions.  By
convention (ISOCAM Consortium), the FF is normalised to one in the
central 144 pixels of the camera. A dark removal is required only for
this FF (see Sect.~\ref{ss:rd}); thus, no precise dark is really
required in the entire reduction procedure. The dark level, even if it
slowly fluctuates, is removed by the beam-switch technique.  The FF is
applied to both the data and the noise cubes ($u$ and $\sigma_u$).

The data are then projected onto a sky map (RA, DEC: epoch 2000) with
a closest pixel method and fine sampled pixels (we use 1.5 arcsecond
pixels on the sky for the 6 arcsecond lens, an integer multiple so
that the method preserves fluxes). A projection of each CAM raster
plane is done by using the raster central position (RA, DEC) and roll
angle (ROLL) from the IIPH raster averaged values.  Corrections are
done for the slightly distorted camera field-of-view.  A gnomonic
projection type is used as with IRAS.  Usually a pixel on the sky has
been measured by different pixels of the camera.  We use an optimised
averaging of these different measurements with a $1/\sigma_u^2$
weighting.  A sky noise map is then also deduced from this optimised
averaging. A scaling is applied to this sky noise map in order to take
into account the correlation of noise in the oversampled sky pixel map
across each camera pixel extension.

\subsection{Detection of point-sources and photometry}\label{ss:psf}

So far, no bias has been introduced against positive or negative (if
any!) flux sources, extended or point sources, except for sources more
extended than the raster step for which the beam-switching reduction
technique is not appropriate. A side effect of 
the method is to leave negative sources of half flux near any positive
source along the raster direction at a raster step distance.  Indeed,
the fitting algorithm of Sect.~\ref{ss:LTD} will find a negative
signal of minus half the central value, on the raster positions that
are adjacent to a strong positive signal position. A more
sophisticated algorithm would be required near the confusion limit.

Point sources are searched for with a top-hat 2 dimensional wavelet.
A 2D Gaussian of fixed width is then fitted (with a simple least
square method) around the candidate position on the sky map in
intensity and position along with a flat background (see a discussion
by Irwin, 1985). For the HDF, we have used an 8 arcsecond (resp. 4)
FWHM Gaussian for the 6 (resp. 3) arcsecond lens.  This routine was
implemented to deal with both the undefined values, which are
scattered around because of the masking applied to the data cube and
its projection onto the sky, and the non-uniform noise maps (affecting
the weight of each pixel in the fit in the optimal $1/\sigma^2$ way).
The final flux of a source is given as the 2D Gaussian integrated
ADUG.  The error on the flux is deduced from the error sky map that
was produced in the previous section.  Fluxes in \microJy are obtained
by dividing fluxes by the integration time per readout
(ADUG/$t_{int}$) and the conversion table in the ISOCAM cookbook
multiplied with an efficiency factor which happens to be unity (the
temporary $\simeq 0.8$ absolute calibration for stabilised point sources
quoted by Cesarsky \etal (1996) has since been revised upward).  The
fluxes are then scaled by a number 1.83 (resp. 1.87) for LW2 (resp.
LW3) that was determined by simulating the ISO PSF, its modulation
by the triple--beam method (that produces negative half-flux ghost
sources on the sides) and the Gaussian fit with a fixed FWHM, in order
to recover the whole camera efficiency.  Fluxes are given at the
nominal wavelength of 6.75 (resp.  15) \micron with an assumed spectral
dependence $F_\nu \propto \nu^{-1}$ (IRAS convention). After the
strongest source is found and its Gaussian flux removed from the map,
one repeats the Gaussian fitting procedure to find the next source. It
is then removed and one iterates the method in order to produce a
catalog of \emph{candidate} sources with position, flux and errors.

\subsection{Detection of slightly extended sources}\label{ss:ses}

The Gaussian fitting allows going to the faintest level of point
source detection but misses part of the flux if the source is
extended. We can also compute a fixed aperture photometric flux in
order to check for possible extensions with different apertures
(although for the particular HDF ISOCAM data we have not done so
because of the small raster steps).  These fluxes are noisier but they
allow a more appropriate measurement for extended sources.  Geometric
parameters can also be deduced following the methods of Jarvis \&
Tyson (1981) and Williams \etal (1996).

\subsection{Reproducibility}\label{ss:rep}

The redundancy factor (the number of times a sky pixel was seen by
different pixels on the camera) is a key factor in deciding the
reliability of sources.  So far we have kept as reliable candidates
those sources which have been covered during at least two raster
positions by unmasked camera pixels. Quality criteria on the
photometric consistency can then be given to each point-source
candidate. For this purpose, we can independently project three
subrasters made out of every third raster values of the $u$ cube onto
the sky. For each subraster, we measure the flux ($F_{1s}$, $F_{2s}$,
$F_{3s}$) and flux uncertainty ($U_{1s}$, $U_{2s}$, $U_{3s}$) of each
source $s$ found in the total map (of flux $F_{ts}$ and noise
$U_{ts}$) at the same position. We define the quality $Q1_s$ of a
source as the highest ranking condition that it meets, according to:

\begin{eqnarray}
\nonumber Q1_s & = 4 {\te if}\ |F_{is}- F_{ts}| \le 2 U_{is}\\
\nonumber      & = 3 {\te if}\ |F_{is}- F_{ts}| \le 3 U_{is}\\
\nonumber      & = 2 {\te if}\ |F_{is}- F_{ts}| \le 4 U_{is}\\
\nonumber      & = 1 {\te if}\  F_{is} \ge 3 U_{is} {\te and\ }  
    F_{ts} \ge 5 U_{ts}\\
     & = 0 {\te otherwise,} \label{eq:Q1}
\end{eqnarray}
where the condition must hold for all the 3 subraster index $i$ (from
1 to 3).  Clearly, the quality from 2 to 4 gives an increasing
confidence in the source reliability.  We added the level of 1 for
strong signal-to-noise sources which can be otherwise dropped because
the flux reproducibility (and not the statistical significance) is
then more difficult to achieve: this source noise happens because of
the ISO jitter, the errors in the projection process and the
undersampling of the PSF. We define a secondary quality $Q2_s$
criterion for low redundancy surveys according to the same logic as
Eq.~\ref{eq:Q1} but keeping only the two flux values with the lowest
noise out of the 3 subraster fluxes.

The reduction of several independent surveys of the same area
(slightly shifted if possible) allows a control of systematics.  The
relative photometric accuracy is generally achieved at the ten percent
level while the absolute photometric accuracy is not better than the
thirty percent level (as deduced from weak calibrated stars and using
the known linearity of the stabilised camera).  It seems that stronger
sources do not have better signal-to-noise (than say 30) because other
errors (the source noise mentioned above is proportional to the
signal) can occur. We found that sources in common in the different
surveys agree well in position (say within a CAM pixel or two).  At present,
the comparison of various ISOCAM datasets with other surveys (in
optical and radio) confirms the astrometric precision of 6 arcsecond
radius at the $2 \sigma$ level.

\section{The HDF data analysis}\label{se:hdf}

\begin{table*}
\caption{\label{ta:hdf2lw2} : HDF-2\_LW2 
list of significant sources at the $3 \sigma$
level found in the raster with TDT no~22401501, 
corrected by -1.90,  -1.80 arcseconds}
\begin{flushleft}
\begin{tiny}
\begin{tabular}{|l|rrr|rrr|rrr|rrr|rrr|rrr|cc|} \hline
 Name &  & $\alpha$           && & $\delta$ &&F1      & U1     & S/N1       & F2     & U2     & S/N2      & F3      & U3    & S/N3       & Ft    & Ut   & S/Nt & Q1 & Q2 \\
   & hr      &mn & sec & deg & '& & $\mu$Jy & $\mu$Jy && $\mu$Jy & $\mu$Jy && $\mu$Jy & $\mu$Jy &&  $\mu$Jy & $\mu$Jy && &  \\
\hline \hline
HDF-2\_LW2\_1&   12& 36& 46.4&   62& 14&  3.4&    91.&     26.&     3.6&       32.&     26.&     1.2&       82.&     26.&     3.1&       69.&     15.&     4.6&    4& 4\\     
HDF-2\_LW2\_2&   12& 36& 48.3&   62& 14& 26.4&    50.&     29.&     1.8&      100.&     27.&     3.6&      133.&     27.&     5.0&       96.&     16.&     6.0&    4& 4\\     
HDF-2\_LW2\_3&   12& 36& 49.7&   62& 13& 46.9&    11.&     26.&     0.4&       79.&     27.&     3.0&       59.&     26.&     2.2&       48.&     15.&     3.2&    4& 4\\     
HDF-2\_LW2\_4&   12& 36& 52.3&   62& 13& 14.9&    25.&     27.&     0.9&       41.&     27.&     1.5&       82.&     26.&     3.1&       49.&     15.&     3.2&    4& 4\\     
\hline
\end{tabular}

\end{tiny}
\end{flushleft}
\end{table*}

\begin{table*}
\caption{\label{ta:hdf4lw2} HDF-4\_LW2 
list of significant sources at the $3 \sigma$
level found in the raster with TDT no~22401305, 
corrected by 0,  0 arcseconds}
\begin{flushleft}
\begin{tiny}
\begin{tabular}{|l|rrr|rrr|rrr|rrr|rrr|rrr|cc|} \hline
 Name &  & $\alpha$           && & $\delta$ &&F1      & U1     & S/N1       & F2     & U2     & S/N2      & F3      & U3    & S/N3       & Ft    & Ut   & S/Nt & Q1 & Q2 \\
   & hr      &mn & sec & deg & '& & $\mu$Jy & $\mu$Jy && $\mu$Jy & $\mu$Jy && $\mu$Jy & $\mu$Jy &&  $\mu$Jy & $\mu$Jy && &  \\
\hline \hline
HDF-4\_LW2\_1&   12& 36& 36.8&   62& 12& 10.8&    62.&     44.&     1.4&      152.&     40.&     3.8&       33.&     43.&     0.8&       85.&     24.&     3.5&    4& 4\\     
HDF-4\_LW2\_2&   12& 36& 37.3&   62& 11& 30.6&   330.&     92.&     3.6&      177.&     77.&     2.3&       43.&     75.&     0.6&      172.&     46.&     3.7&    4& 4\\     
\hline
\end{tabular}

\end{tiny}
\end{flushleft}
\end{table*}

\begin{table*}
\caption{\label{ta:hdf2lw3} : HDF-2\_LW3 
list of significant sources at the $3 \sigma$
level found in the raster with TDT no~22701702,
corrected by -1.20,  -4.10 arcseconds}
\begin{flushleft}
\begin{tiny}
\begin{tabular}{|l|rrr|rrr|rrr|rrr|rrr|rrr|cc|} \hline
 Name &  & $\alpha$           && & $\delta$ &&F1      & U1     & S/N1       & F2     & U2     & S/N2      & F3      & U3    & S/N3       & Ft    & Ut   & S/Nt & Q1 & Q2 \\
   & hr      &mn & sec & deg & '& & $\mu$Jy & $\mu$Jy && $\mu$Jy & $\mu$Jy && $\mu$Jy & $\mu$Jy &&  $\mu$Jy & $\mu$Jy && &  \\
\hline \hline
HDF-2\_LW3\_1&   12& 36& 34.2&   62& 12& 30.6&   337.&    147.&     2.3&      209.&    133.&     1.6&      442.&    154.&     2.9&      322.&     83.&     3.9&    4& 4\\     
HDF-2\_LW3\_2&   12& 36& 36.2&   62& 13& 39.0&   182.&     75.&     2.4&      288.&     74.&     3.9&      289.&     73.&     4.0&      254.&     43.&     6.0&    4& 4\\     
HDF-2\_LW3\_3&   12& 36& 39.2&   62& 14& 23.4&    53.&     67.&     0.8&      144.&     70.&     2.1&      183.&     75.&     2.4&      123.&     41.&     3.0&    4& 4\\     
HDF-2\_LW3\_4&   12& 36& 39.6&   62& 12& 42.9&   217.&     65.&     3.4&      231.&     65.&     3.6&      403.&     65.&     6.2&      284.&     37.&     7.6&    4& 4\\     
HDF-2\_LW3\_5&   12& 36& 43.9&   62& 12& 44.5&   161.&     55.&     3.0&      138.&     56.&     2.5&      264.&     56.&     4.7&      188.&     32.&     5.8&    4& 4\\     
HDF-2\_LW3\_6&   12& 36& 46.3&   62& 14&  3.9&   153.&     55.&     2.8&       49.&     54.&     0.9&      166.&     54.&     3.1&      122.&     31.&     3.9&    4& 4\\     
HDF-2\_LW3\_7&   12& 36& 46.4&   62& 15& 30.2&    81.&     80.&     1.0&      383.&     73.&     5.2&      291.&     74.&     3.9&      256.&     43.&     5.9&    3& 4\\     
HDF-2\_LW3\_8&   12& 36& 46.5&   62& 14& 48.3&    87.&     54.&     1.6&      114.&     53.&     2.2&      170.&     58.&     2.9&      122.&     32.&     3.9&    4& 4\\     
HDF-2\_LW3\_9&   12& 36& 48.4&   62& 14& 26.4&   308.&     53.&     5.8&      241.&     54.&     4.4&      219.&     55.&     4.0&      256.&     31.&     8.2&    4& 4\\     
HDF-2\_LW3\_10&   12& 36& 49.3&   62& 14&  5.5&    52.&     56.&     0.9&      167.&     57.&     2.9&       73.&     54.&     1.3&       98.&     32.&     3.0&    4& 4\\     
HDF-2\_LW3\_11&   12& 36& 49.7&   62& 13& 13.0&   259.&     53.&     4.9&      260.&     51.&     5.1&      243.&     52.&     4.7&      254.&     30.&     8.5&    4& 4\\     
HDF-2\_LW3\_12&   12& 36& 51.1&   62& 13& 17.0&   174.&     53.&     3.3&       67.&     51.&     1.3&       44.&     54.&     0.8&       95.&     30.&     3.1&    4& 4\\     
HDF-2\_LW3\_13&   12& 36& 51.8&   62& 13& 54.1&   142.&     54.&     2.6&      238.&     53.&     4.5&      112.&     54.&     2.1&      165.&     31.&     5.3&    4& 4\\     
HDF-2\_LW3\_14&   12& 36& 53.9&   62& 12& 53.6&   109.&     52.&     2.1&      183.&     53.&     3.4&      163.&     55.&     2.9&      151.&     31.&     4.9&    4& 4\\     
HDF-2\_LW3\_15&   12& 36& 57.9&   62& 14& 56.5&    91.&     60.&     1.5&      147.&     66.&     2.2&      266.&     62.&     4.3&      165.&     36.&     4.6&    4& 4\\     
HDF-2\_LW3\_16&   12& 36& 60.0&   62& 14& 53.2&   276.&     62.&     4.5&      352.&     65.&     5.4&      268.&     65.&     4.2&      299.&     37.&     8.1&    4& 4\\     
HDF-2\_LW3\_17&   12& 37&  2.5&   62& 14&  4.3&   196.&     61.&     3.2&      188.&     67.&     2.8&      187.&     63.&     3.0&      190.&     37.&     5.2&    4& 4\\     
HDF-2\_LW3\_18&   12& 37&  4.9&   62& 14& 31.2&   115.&     76.&     1.5&      181.&     79.&     2.3&      120.&     78.&     1.5&      137.&     45.&     3.1&    4& 4\\     
\hline
\end{tabular}

\end{tiny}
\end{flushleft}
\end{table*}

\begin{figure*}[htbp]    
  \includegraphics[width=\textwidth, angle=90, origin=br]
    {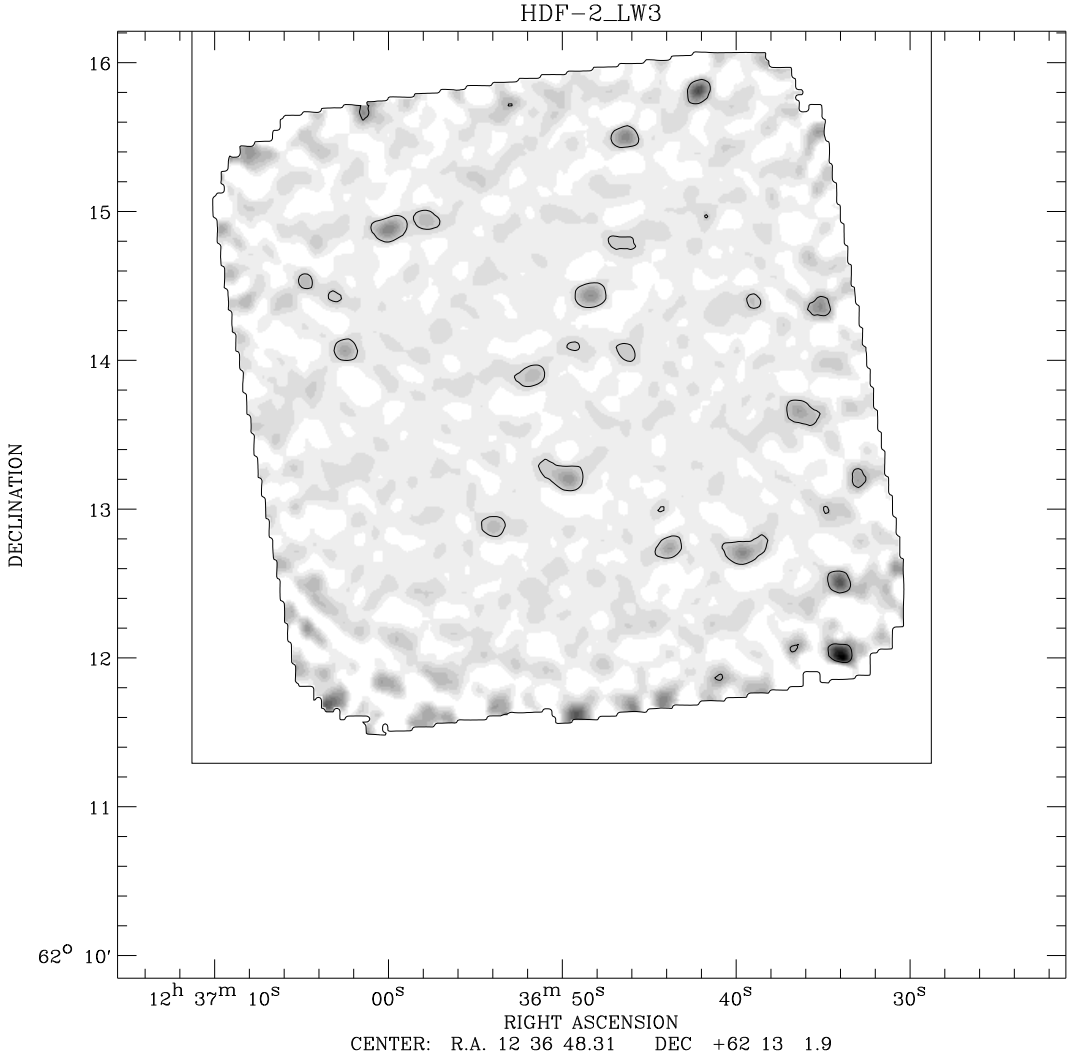}
  \caption{LW3 grey scale map of the HDF-2 raster. The contour is at
about the 3~$\sigma$ level. Black corresponds to about 300\microJy per
beam, white to 0. Following maps are done on the same scale.}
  \label{fi:2lw3}
\end{figure*}
\begin{figure*}[htbp]    
  \includegraphics[width=\textwidth, angle=90, origin=br]
   {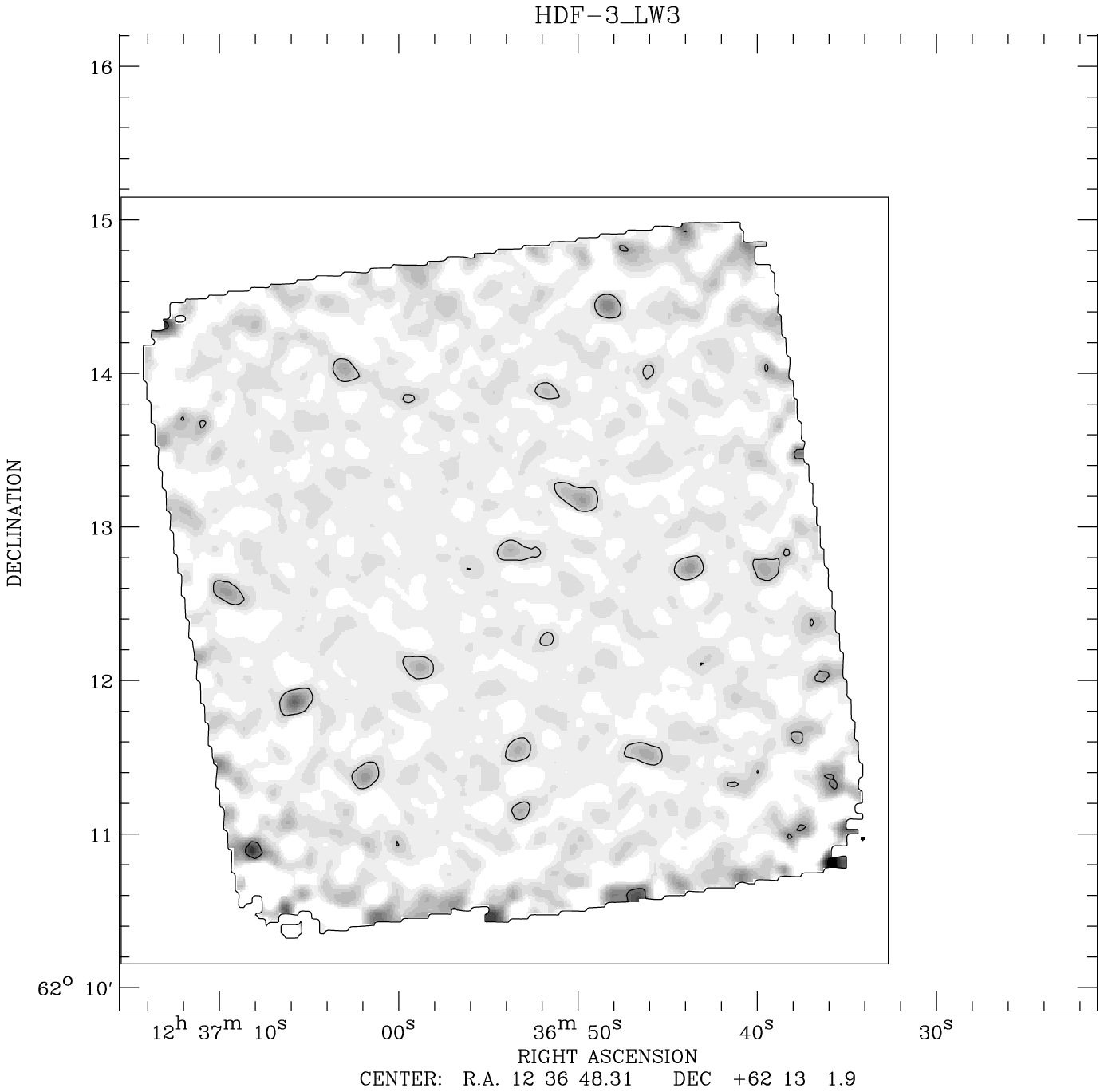}
  \caption{HDF LW3 grey scale map of the HDF-3 raster. Contour
and greyscale are identical to the previous figure.}
  \label{fi:3lw3}
\end{figure*}
\begin{figure*}[htbp]    
  \includegraphics[width=\textwidth, angle=90, origin=br]
     {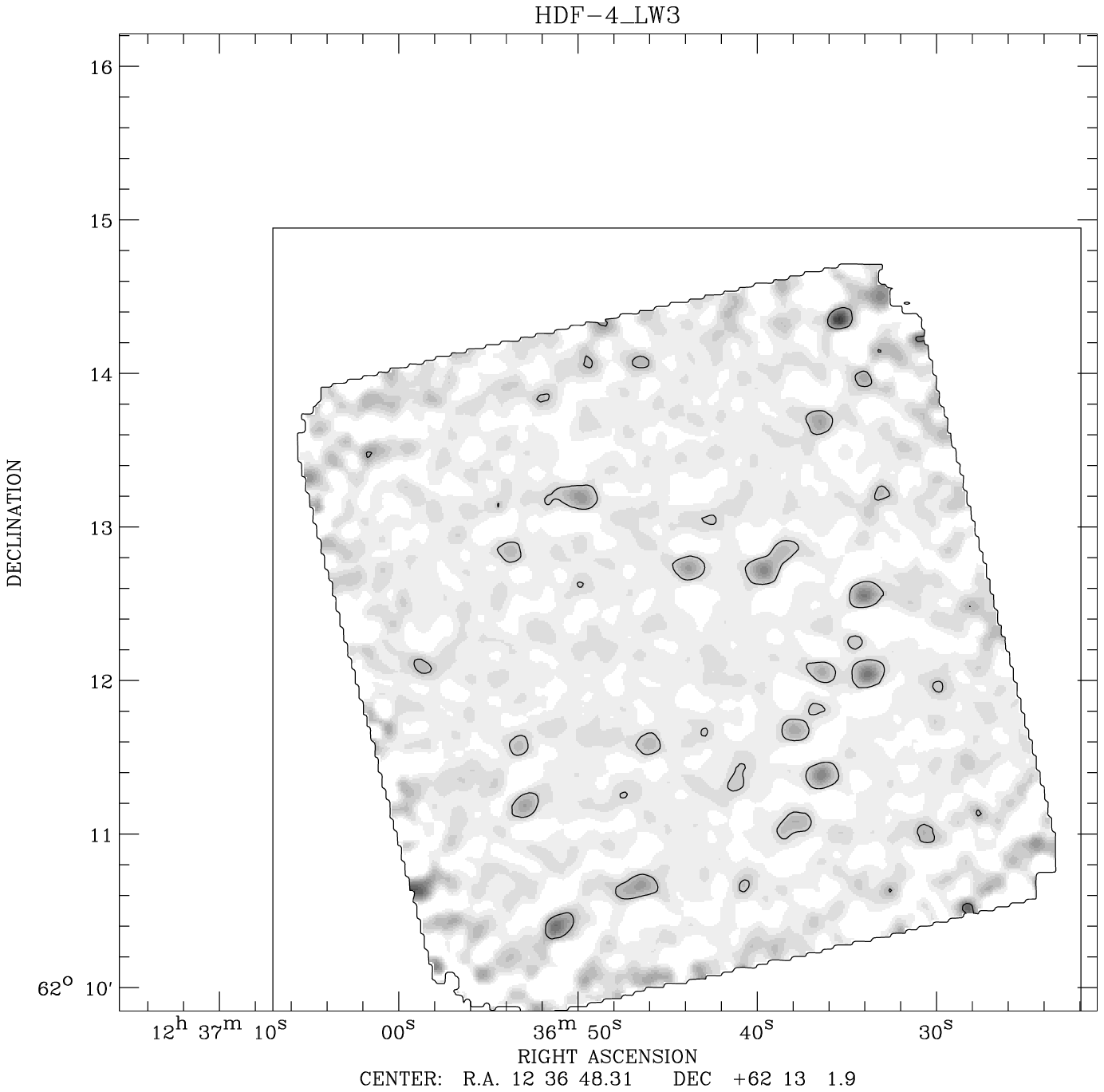}
  \caption{HDF LW3 grey scale map of the HDF-4 raster. Contour
and greyscale are identical to the previous figure.}
  \label{fi:4lw3}
\end{figure*}

The data in the public domain on the Hubble Deep Field in the LW2
(6.7\micron) and LW3 (15\micron) ISOCAM-LW bands are analysed here
with the method outlined above.  Early results are given by
Rowan--Robinson \etal (1997). The data were taken in three rasters for
each wavelength centered on the optical centre of the Hubble Space
Telescope WFPC quadrants number 2, 3 and 4.  The 3 arcsecond lens was
used for LW2 and the 6 arcsecond lens for LW3. The LW3 rasters are
therefore largely overlapping.  Adjacent raster positions are
separated by resp. 5 and 9 arcseconds, roughly a pixel and a half
apart. The resulting triple beam-switch method that we apply may miss
a fraction of the flux (although we have tried to evaluate it in
Sect.~\ref{ss:psf}), in that particular case, especially if there
are extended objects. Each raster is made of 8 by 8 positions aligned
with the camera axes (see Fig.~\ref{fi:iiph}). Each LW2 (resp. LW3)
raster position is made of resp.  9 to 10 (19 to 20) readouts of 72
(resp. 36) CAMTU (10 and 5 s resp.)  integration times.
Tables~\ref{ta:hdf2lw2} and ~\ref{ta:hdf4lw2} and
Tables~\ref{ta:hdf2lw3}, \ref{ta:hdf3lw3}, \&~\ref{ta:hdf4lw3} give
the details of potential sources with a quality different from 0
(either $Q1$ or $Q2$) for each ISO raster (no significant source was
found with LW2 in the third HDF quadrant).  The source positions for
each raster are corrected by an absolute astrometric offset. This
offset was determined from the sources which are detected in several
rasters, the optical identification for the primary astrometric
corrector is a 19.5 magnitude object (with a redshift of 0.139, Cowie
\etal 1998) situated at (J2000) RA= 12~36~48.33 and Dec=~62~14~26.4 at
the upper part of the HDF, which is detected at both wavelengths.
Note that we could not coadd LW2 rasters (they do not overlap much
anyway) because of the lack of common sources or well identified
sources.  Figure~\ref{fi:2lw3}, \ref{fi:3lw3} and \ref{fi:4lw3} show the
3 independent LW3 rasters with intensity in greyscale and 3 $\sigma$
contours.  One can notice at once the large consistency between the 3
independent datasets in the overlapping area.  Figure~\ref{fi:vlw3} shows
an f814 HST image of the HDF (Williams \etal 1996) superposed by a 3
$\sigma$ contour of ISOCAM after shifting and optimally coadding the three
individual LW3 rasters.  Table~\ref{ta:hdfalllw3} presents the final
list of objects detected in the LW3 coadded map.  Figure~\ref{fi:klw3}
shows a K-band IRIM image of the HDF (Dickinson \etal 1997) superposed
by a 3 $\sigma$ contour of the same ISOCAM LW3 total map. It is clear
that the sources in the final catalog (table~\ref{ta:hdfalllw3}) that
are not in one or two of the raster tables (\ref{ta:hdf2lw3},
\ref{ta:hdf3lw3}, \&~\ref{ta:hdf4lw3}) are simply outside or at the
edge of the corresponding observed area.
\begin{table*}
\caption{\label{ta:hdf3lw3} : HDF-3\_LW3 
list of significant sources at the $3 \sigma$
level found in the raster with TDT no~22701604,
corrected by -1.20,  -7.10 arcseconds}
\begin{flushleft}
\begin{tiny}
\begin{tabular}{|l|rrr|rrr|rrr|rrr|rrr|rrr|cc|} \hline
 Name &  & $\alpha$           && & $\delta$ &&F1      & U1     & S/N1       & F2     & U2     & S/N2      & F3      & U3    & S/N3       & Ft    & Ut   & S/Nt & Q1 & Q2 \\
   & hr      &mn & sec & deg & '& & $\mu$Jy & $\mu$Jy && $\mu$Jy & $\mu$Jy && $\mu$Jy & $\mu$Jy &&  $\mu$Jy & $\mu$Jy && &  \\
\hline \hline
HDF-3\_LW3\_1&   12& 36& 39.4&   62& 12& 44.0&   349.&     83.&     4.2&      137.&     81.&     1.7&      213.&     83.&     2.6&      232.&     47.&     4.9&    4& 4\\     
HDF-3\_LW3\_2&   12& 36& 42.9&   62& 12&  8.2&   152.&     56.&     2.7&      133.&     59.&     2.2&       25.&     60.&     0.4&      106.&     34.&     3.1&    4& 4\\     
HDF-3\_LW3\_3&   12& 36& 43.8&   62& 12& 44.3&   241.&     56.&     4.3&      199.&     59.&     3.3&      224.&     60.&     3.7&      222.&     34.&     6.6&    4& 4\\     
HDF-3\_LW3\_4&   12& 36& 46.2&   62& 11& 31.8&   157.&     63.&     2.5&      276.&     59.&     4.7&      167.&     60.&     2.8&      200.&     35.&     5.7&    4& 4\\     
HDF-3\_LW3\_5&   12& 36& 48.3&   62& 14& 26.4&   431.&     81.&     5.3&      167.&     77.&     2.2&      253.&     83.&     3.0&      279.&     46.&     6.1&    4& 4\\     
HDF-3\_LW3\_6&   12& 36& 49.8&   62& 13& 11.7&   177.&     53.&     3.3&      184.&     54.&     3.4&      344.&     53.&     6.5&      236.&     31.&     7.7&    3& 3\\     
HDF-3\_LW3\_7&   12& 36& 51.7&   62& 13& 53.2&   140.&     57.&     2.5&       76.&     60.&     1.3&      136.&     57.&     2.4&      119.&     33.&     3.6&    4& 4\\     
HDF-3\_LW3\_8&   12& 36& 53.2&   62& 11&  9.3&    95.&     79.&     1.2&      211.&     75.&     2.8&      150.&     76.&     2.0&      147.&     45.&     3.3&    4& 4\\     
HDF-3\_LW3\_9&   12& 36& 53.3&   62& 11& 33.1&   185.&     62.&     3.0&      218.&     57.&     3.8&      132.&     56.&     2.4&      180.&     34.&     5.4&    4& 4\\     
HDF-3\_LW3\_10&   12& 36& 53.6&   62& 12& 50.9&   101.&     53.&     1.9&      137.&     55.&     2.5&      267.&     56.&     4.8&      167.&     32.&     5.3&    4& 4\\     
HDF-3\_LW3\_11&   12& 36& 58.9&   62& 12&  5.2&   205.&     52.&     3.9&      200.&     55.&     3.6&      157.&     53.&     3.0&      188.&     31.&     6.1&    4& 4\\     
HDF-3\_LW3\_12&   12& 37&  1.9&   62& 11& 23.1&   149.&     73.&     2.1&      434.&     75.&     5.8&      147.&     66.&     2.2&      239.&     41.&     5.8&    3& 4\\     
HDF-3\_LW3\_13&   12& 37&  3.1&   62& 14&  1.2&   112.&     71.&     1.6&      281.&     70.&     4.0&      188.&     71.&     2.6&      203.&     41.&     5.0&    4& 4\\     
HDF-3\_LW3\_14&   12& 37&  5.8&   62& 11& 51.6&   386.&     75.&     5.2&      483.&     78.&     6.2&      234.&     76.&     3.1&      367.&     44.&     8.4&    4& 4\\     
HDF-3\_LW3\_15&   12& 37&  9.4&   62& 12& 34.2&   199.&    100.&     2.0&      207.&    100.&     2.1&      292.&    102.&     2.9&      229.&     58.&     4.0&    4& 4\\     
\hline
\end{tabular}

\end{tiny}
\end{flushleft}
\end{table*}
\begin{table*}
\caption{\label{ta:hdf4lw3} HDF-4\_LW3 list 
of significant sources at the $3 \sigma$
level found in the raster with TDT no~22202606, 
corrected by -1.20,  -1.30 arcseconds}
\begin{flushleft}
\begin{tiny}
\begin{tabular}{|l|rrr|rrr|rrr|rrr|rrr|rrr|cc|} \hline
 Name &  & $\alpha$           && & $\delta$ &&F1      & U1     & S/N1       & F2     & U2     & S/N2      & F3      & U3    & S/N3       & Ft    & Ut   & S/Nt & Q1 & Q2 \\
   & hr      &mn & sec & deg & '& & $\mu$Jy & $\mu$Jy && $\mu$Jy & $\mu$Jy && $\mu$Jy & $\mu$Jy &&  $\mu$Jy & $\mu$Jy && &  \\
\hline \hline
HDF-4\_LW3\_1&   12& 36& 29.9&   62& 11& 58.4&   126.&     74.&     1.7&      227.&     75.&     3.0&       31.&     75.&     0.4&      130.&     43.&     3.0&    4& 4\\     
HDF-4\_LW3\_2&   12& 36& 30.6&   62& 11&  0.5&   211.&     88.&     2.4&       33.&     90.&     0.4&      308.&     89.&     3.4&      184.&     51.&     3.6&    4& 4\\     
HDF-4\_LW3\_3&   12& 36& 33.1&   62& 13& 12.4&     5.&     78.&     0.1&      231.&     79.&     2.9&      209.&     77.&     2.7&      146.&     45.&     3.3&    4& 4\\     
HDF-4\_LW3\_4&   12& 36& 33.8&   62& 12&  3.0&   301.&     54.&     5.6&      287.&     55.&     5.2&      426.&     56.&     7.6&      337.&     32.&    10.6&    4& 4\\     
HDF-4\_LW3\_5&   12& 36& 33.9&   62& 12& 33.8&   383.&     62.&     6.2&      283.&     63.&     4.5&      246.&     59.&     4.2&      305.&     35.&     8.7&    4& 4\\     
HDF-4\_LW3\_6&   12& 36& 34.1&   62& 13& 57.6&   127.&     97.&     1.3&      161.&    107.&     1.5&      257.&    101.&     2.6&      179.&     58.&     3.1&    4& 4\\     
HDF-4\_LW3\_7&   12& 36& 36.3&   62& 12&  3.2&   241.&     53.&     4.6&      136.&     52.&     2.6&      168.&     54.&     3.1&      181.&     30.&     6.0&    4& 4\\     
HDF-4\_LW3\_8&   12& 36& 36.4&   62& 11& 23.0&   230.&     53.&     4.4&      332.&     53.&     6.2&      291.&     54.&     5.4&      285.&     31.&     9.3&    4& 4\\     
HDF-4\_LW3\_9&   12& 36& 36.5&   62& 13& 41.0&   194.&     65.&     3.0&      186.&     68.&     2.8&      269.&     66.&     4.1&      216.&     38.&     5.7&    4& 4\\     
HDF-4\_LW3\_10&   12& 36& 37.9&   62& 11&  4.2&   179.&     65.&     2.8&      189.&     62.&     3.1&      231.&     60.&     3.9&      201.&     36.&     5.6&    4& 4\\     
HDF-4\_LW3\_11&   12& 36& 37.9&   62& 11& 41.3&   235.&     53.&     4.4&      173.&     56.&     3.1&      163.&     54.&     3.0&      190.&     31.&     6.1&    4& 4\\     
HDF-4\_LW3\_12&   12& 36& 38.4&   62& 12& 51.2&   167.&     53.&     3.2&      125.&     53.&     2.4&      172.&     54.&     3.2&      153.&     31.&     5.0&    4& 4\\     
HDF-4\_LW3\_13&   12& 36& 39.6&   62& 12& 43.3&   291.&     53.&     5.5&      388.&     53.&     7.3&      231.&     55.&     4.2&      306.&     31.&     9.9&    4& 4\\     
HDF-4\_LW3\_14&   12& 36& 41.1&   62& 11& 21.3&   134.&     53.&     2.5&      141.&     54.&     2.6&      131.&     55.&     2.4&      135.&     31.&     4.3&    4& 4\\     
HDF-4\_LW3\_15&   12& 36& 42.7&   62& 13&  2.8&    32.&     56.&     0.6&      161.&     57.&     2.8&      111.&     55.&     2.0&      101.&     32.&     3.1&    4& 4\\     
HDF-4\_LW3\_16&   12& 36& 43.8&   62& 12& 44.1&   228.&     54.&     4.3&      219.&     54.&     4.0&      200.&     57.&     3.5&      215.&     32.&     6.8&    4& 4\\     
HDF-4\_LW3\_17&   12& 36& 46.1&   62& 11& 35.5&   166.&     54.&     3.1&      182.&     55.&     3.3&       71.&     55.&     1.3&      140.&     32.&     4.4&    4& 4\\     
HDF-4\_LW3\_18&   12& 36& 46.6&   62& 10& 39.8&   293.&     73.&     4.0&      369.&     74.&     5.0&      167.&     74.&     2.3&      277.&     42.&     6.6&    4& 4\\     
HDF-4\_LW3\_19&   12& 36& 49.9&   62& 13& 12.1&   291.&     56.&     5.2&      240.&     54.&     4.4&      256.&     57.&     4.5&      262.&     32.&     8.2&    4& 4\\     
HDF-4\_LW3\_20&   12& 36& 51.1&   62& 10& 24.0&   385.&    122.&     3.2&      371.&    127.&     2.9&      374.&    114.&     3.3&      376.&     70.&     5.4&    4& 4\\     
HDF-4\_LW3\_21&   12& 36& 52.9&   62& 11& 11.2&   131.&     68.&     1.9&      241.&     66.&     3.7&      231.&     72.&     3.2&      203.&     39.&     5.1&    4& 4\\     
HDF-4\_LW3\_22&   12& 36& 53.3&   62& 11& 35.2&   108.&     61.&     1.8&       97.&     64.&     1.5&      174.&     63.&     2.8&      127.&     36.&     3.5&    4& 4\\     
HDF-4\_LW3\_23&   12& 36& 53.8&   62& 12& 50.3&   170.&     54.&     3.2&      152.&     59.&     2.6&      102.&     55.&     1.9&      140.&     32.&     4.4&    4& 4\\     
\hline
\end{tabular}

\end{tiny}
\end{flushleft}
\end{table*}

\begin{table*}
\caption{\label{ta:hdfalllw3} List of significant sources at the $3 \sigma$
level found in final combined LW3 map. Positions in J2000 system.
Last column indicates likely
associations with an LW2 source
(Table~\ref{ta:hdf2lw2}\&~\ref{ta:hdf4lw2}), with the radio source
catalog by Richards \etal (1998) and with a visible-IR source with redshift
from the active catalogs by Cowie \etal (1998).}
\begin{flushleft}
\begin{tiny}
\begin{tabular}{|l|rrr|rrr|rrr|rrr|rrr|rrr|cc|l|} \hline
 Name &  & $\alpha$           && & $\delta$ &&F1      & U1     & S/N1
   & F2     & U2     & S/N2      & F3      & U3    & S/N3       & Ft
   & Ut   & S/Nt & Q1 & Q2 & LW2, R, z  \\
   & hr      &mn & sec & deg & '& & $\mu$Jy & $\mu$Jy && $\mu$Jy & $\mu$Jy && $\mu$Jy & $\mu$Jy &&  $\mu$Jy & $\mu$Jy && & & \\
\hline \hline
HDF\_ALL\_LW3\_1&   12& 36& 30.6&   62& 11&  0.5&   293.&     93.&
   3.2&      234.&     89.&     2.6&        6.&     95.&     0.1&
   183.&     53.&     3.4&    4& 4 & R\\     
HDF\_ALL\_LW3\_2&   12& 36& 33.0&   62& 13& 12.6&   217.&     65.&
   3.4&      108.&     67.&     1.6&      172.&     65.&     2.6&
   165.&     38.&     4.4&    4& 4 & \\     
HDF\_ALL\_LW3\_3&   12& 36& 33.8&   62& 12&  2.9&   411.&     54.&
   7.7&      332.&     53.&     6.2&      300.&     55.&     5.5&
   348.&     31.&    11.2&    4& 4 & \\     
HDF\_ALL\_LW3\_4&   12& 36& 34.0&   62& 12& 33.0&   263.&     53.&
   4.9&      337.&     55.&     6.1&      342.&     58.&     5.9&
   314.&     32.&     9.8&    4& 4 & R, 1.219\\     
HDF\_ALL\_LW3\_5&   12& 36& 34.2&   62& 13& 58.2&   164.&     73.&     2.2&       83.&     73.&     1.1&      207.&     78.&     2.6&      145.&     43.&     3.4&    4& 4 & \\     
\hline
HDF\_ALL\_LW3\_6&   12& 36& 35.3&   62& 14& 22.1&   258.&     87.&
   3.0&      207.&     93.&     2.2&      580.&     96.&     6.0&
   338.&     53.&     6.4&    3& 4 & R\\     
HDF\_ALL\_LW3\_7&   12& 36& 36.3&   62& 11& 22.9&   263.&     53.&     5.0&      249.&     50.&     5.0&      319.&     52.&     6.1&      278.&     30.&     9.3&    4& 4 & \\     
HDF\_ALL\_LW3\_8&   12& 36& 36.3&   62& 12& 22.4&    55.&     45.&     1.2&       51.&     43.&     1.2&      164.&     46.&     3.6&       88.&     26.&     3.4&    4& 4 & \\     
HDF\_ALL\_LW3\_9&   12& 36& 36.4&   62& 12&  3.3&   139.&     48.&     2.9&      218.&     46.&     4.7&      159.&     46.&     3.4&      174.&     27.&     6.4&    4& 4 & \\     
HDF\_ALL\_LW3\_10&   12& 36& 36.4&   62& 13& 40.2&   217.&     49.&
   4.4&      248.&     49.&     5.1&      209.&     49.&     4.2&
   225.&     28.&     7.9&    4& 4 & 0.556\\     
\hline
HDF\_ALL\_LW3\_11&   12& 36& 37.9&   62& 11& 40.8&   162.&     47.&
   3.4&      228.&     47.&     4.8&      167.&     51.&     3.3&
   187.&     28.&     6.7&    4& 4 & LW2, R, 0.078\\     
HDF\_ALL\_LW3\_12&   12& 36& 37.9&   62& 11&  3.4&   274.&     64.&     4.3&      190.&     63.&     3.0&      172.&     59.&     2.9&      215.&     36.&     6.0&    4& 4 & \\     
HDF\_ALL\_LW3\_13&   12& 36& 38.3&   62& 12& 50.8&    82.&     38.&
   2.1&      153.&     38.&     4.0&      141.&     39.&     3.7&
   125.&     22.&     5.7&    4& 4 & R\\     
HDF\_ALL\_LW3\_14&   12& 36& 39.0&   62& 14& 22.1&    55.&     58.&     1.0&      108.&     57.&     1.9&      177.&     60.&     2.9&      110.&     34.&     3.3&    4& 4 & \\     
HDF\_ALL\_LW3\_15&   12& 36& 39.6&   62& 12& 43.3&   213.&     37.&     5.7&      282.&     37.&     7.7&      330.&     36.&     9.1&      275.&     21.&    13.0&    4& 4 & \\     
\hline
HDF\_ALL\_LW3\_16&   12& 36& 40.7&   62& 10& 41.0&   204.&     82.&     2.5&      166.&     74.&     2.3&      107.&     89.&     1.2&      161.&     48.&     3.4&    4& 4 & \\     
HDF\_ALL\_LW3\_17&   12& 36& 41.1&   62& 11& 20.3&   148.&     48.&     3.1&       83.&     46.&     1.8&      177.&     47.&     3.8&      135.&     27.&     5.0&    4& 4 & \\     
HDF\_ALL\_LW3\_18&   12& 36& 42.1&   62& 15& 48.5&   651.&    113.&     5.8&      156.&    120.&     1.3&      331.&    122.&     2.7&      397.&     68.&     5.8&    3& 3 & \\     
HDF\_ALL\_LW3\_19&   12& 36& 43.8&   62& 12& 44.4&   197.&     33.&
   6.0&      211.&     32.&     6.7&      231.&     32.&     7.1&
   213.&     19.&    11.4&    4& 4 & R, 0.557\\     
HDF\_ALL\_LW3\_20&   12& 36& 46.2&   62& 11& 33.5&   117.&     40.&     2.9&      133.&     40.&     3.3&      199.&     40.&     5.0&      149.&     23.&     6.4&    4& 4 & \\     
\hline
HDF\_ALL\_LW3\_21&   12& 36& 46.3&   62& 14&  3.2&    99.&     38.&
   2.6&      114.&     38.&     3.0&      121.&     38.&     3.1&
   111.&     22.&     5.0&    4& 4 & LW2, R, 0.960\\     
HDF\_ALL\_LW3\_22&   12& 36& 46.4&   62& 15& 30.2&    72.&     80.&     0.9&      408.&     73.&     5.6&      292.&     74.&     3.9&      264.&     44.&     6.1&    3& 4 & \\     
HDF\_ALL\_LW3\_23&   12& 36& 46.7&   62& 10& 38.9&   191.&     69.&     2.8&      273.&     73.&     3.7&      344.&     71.&     4.8&      267.&     41.&     6.5&    4& 4 & \\     
HDF\_ALL\_LW3\_24&   12& 36& 46.8&   62& 14& 47.5&   107.&     52.&
   2.1&      140.&     50.&     2.8&      147.&     53.&     2.8&
   130.&     30.&     4.4&    4& 4 & R\\     
HDF\_ALL\_LW3\_25&   12& 36& 48.4&   62& 14& 26.4&   291.&     45.&
   6.5&      298.&     45.&     6.6&      205.&     45.&     4.6&
   265.&     26.&    10.3&    4& 4 & LW2, R, 0.139\\     
\hline
HDF\_ALL\_LW3\_26&   12& 36& 49.3&   62& 14&  5.6&    57.&     38.&
   1.5&      100.&     38.&     2.6&      115.&     38.&     3.0&
   91.&     22.&     4.2&    4& 4 & 0.751\\     
HDF\_ALL\_LW3\_27&   12& 36& 49.8&   62& 13& 12.3&   281.&     31.&
   9.0&      247.&     31.&     8.1&      222.&     31.&     7.2&
   250.&     18.&    14.0&    4& 4 & R, 0.475\\     
HDF\_ALL\_LW3\_28&   12& 36& 51.1&   62& 10& 23.9&   332.&    114.&     2.9&      374.&    123.&     3.0&      409.&    126.&     3.2&      367.&     70.&     5.2&    4& 4 & \\     
HDF\_ALL\_LW3\_29&   12& 36& 51.2&   62& 13& 14.7&   140.&     32.&
   4.3&       26.&     32.&     0.8&       53.&     33.&     1.6&
   73.&     19.&     3.9&    3& 3 & LW2, R, 0.475\\     
HDF\_ALL\_LW3\_30&   12& 36& 51.8&   62& 13& 53.5&   126.&     35.&
   3.6&      159.&     35.&     4.5&      120.&     36.&     3.3&
   135.&     21.&     6.5&    4& 4 & R, 0.557\\     
\hline
HDF\_ALL\_LW3\_31&   12& 36& 53.0&   62& 11& 10.5&   183.&     52.&     3.5&       98.&     50.&     2.0&      231.&     49.&     4.7&      170.&     29.&     5.9&    4& 4 & \\     
HDF\_ALL\_LW3\_32&   12& 36& 53.3&   62& 11& 34.0&   149.&     41.&
   3.6&      149.&     42.&     3.5&      169.&     42.&     4.1&
   155.&     24.&     6.4&    4& 4 & R\\     
HDF\_ALL\_LW3\_33&   12& 36& 53.8&   62& 12& 51.5&   154.&     31.&     4.9&      133.&     31.&     4.3&      144.&     33.&     4.4&      143.&     18.&     7.9&    4& 4 & 0.642\\     
HDF\_ALL\_LW3\_34&   12& 36& 57.9&   62& 14& 56.6&    97.&     60.&
   1.6&      156.&     65.&     2.4&      260.&     62.&     4.2&
   170.&     36.&     4.7&    4& 4 & R\\     
HDF\_ALL\_LW3\_35&   12& 36& 58.9&   62& 12&  5.4&   138.&     42.&     3.3&      103.&     44.&     2.4&      225.&     43.&     5.3&      156.&     25.&     6.3&    4& 4 & \\     
\hline
HDF\_ALL\_LW3\_36&   12& 37&  0.0&   62& 14& 53.3&   273.&     62.&
   4.4&      356.&     66.&     5.4&      280.&     65.&     4.3&
   301.&     37.&     8.1&    4& 4 & 0.761\\     
HDF\_ALL\_LW3\_37&   12& 37&  1.9&   62& 11& 23.0&   146.&     67.&     2.2&      164.&     73.&     2.3&      420.&     76.&     5.6&      235.&     41.&     5.7&    3& 4 & \\     
HDF\_ALL\_LW3\_38&   12& 37&  2.8&   62& 14&  2.6&   148.&     48.&     3.1&      128.&     49.&     2.6&      228.&     48.&     4.8&      167.&     28.&     6.1&    4& 4 & \\     
HDF\_ALL\_LW3\_39&   12& 37&  5.9&   62& 11& 51.7&   214.&     77.&     2.8&      385.&     75.&     5.1&      467.&     78.&     6.0&      359.&     44.&     8.1&    4& 4 & \\     
HDF\_ALL\_LW3\_40&   12& 37&  8.3&   62& 10& 53.9&   618.&    266.&
   2.3&      367.&    294.&     1.2&      410.&    273.&     1.5&
   501.&    160.&     3.1&    4& 4 & R\\     
\hline
HDF\_ALL\_LW3\_41&   12& 37&  9.6&   62& 12& 34.6&   372.&     96.&
   3.9&      206.&     98.&     2.1&      234.&     90.&     2.6&
   267.&     54.&     4.9&    4& 4 & star\\     
\hline
\end{tabular}

\end{tiny}
\end{flushleft}
\end{table*}

\section{Discussion}\label{se:disc}

\begin{figure*}[htbp]    
  \includegraphics[width=\textwidth, angle=90, origin=br]
  {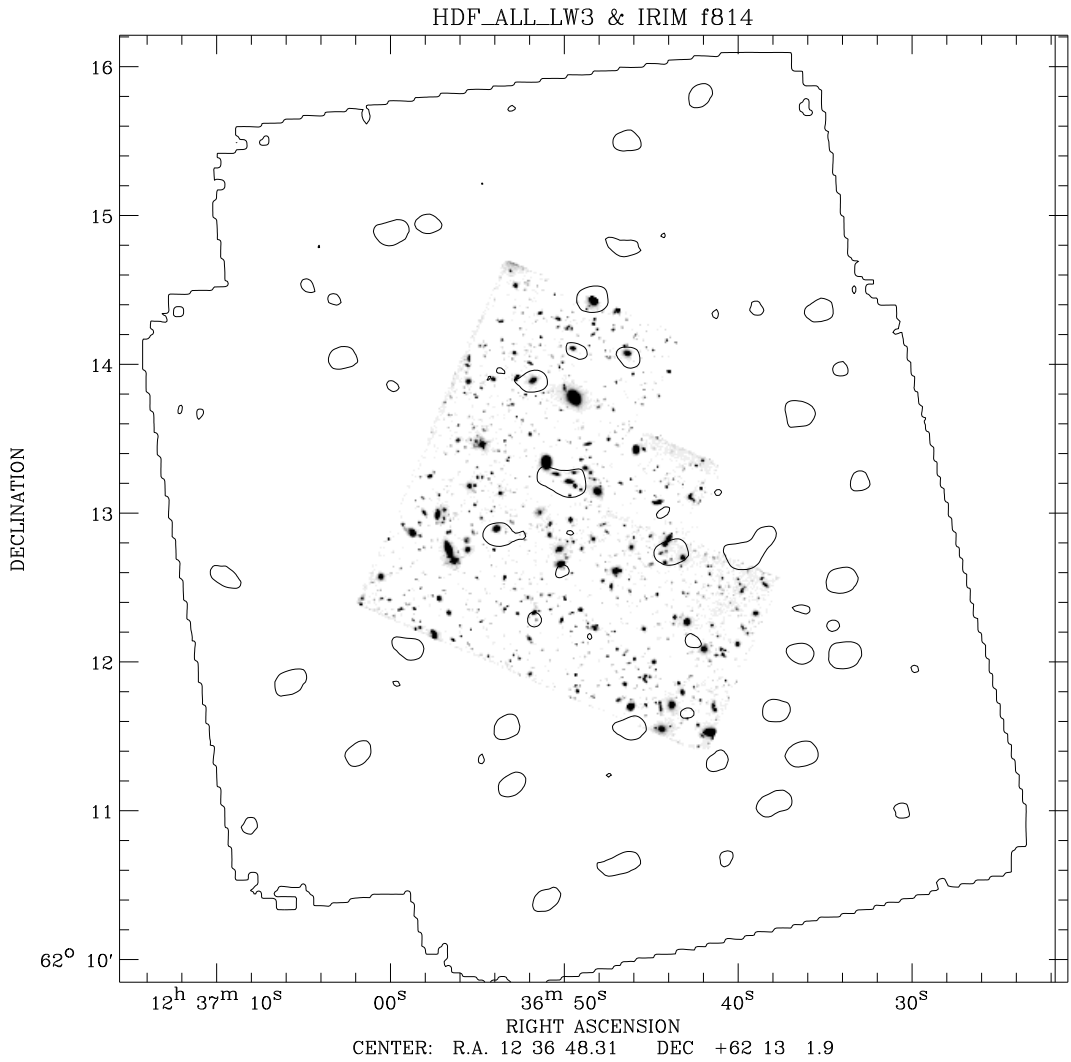}
  \caption{HDF optical f814 map (grey scale) overlaid with
the LW3 contour of the 3 coadded rasters. The contour is at
about the 3~$\sigma$ level. The outside contour 
(as defined by a minimum redundancy of 2) delimits the region
where the source searching algorithm was applied.}
  \label{fi:vlw3}
\end{figure*}
\begin{figure*}[htbp]    
  \includegraphics[width=\textwidth, angle=90, origin=br]
    {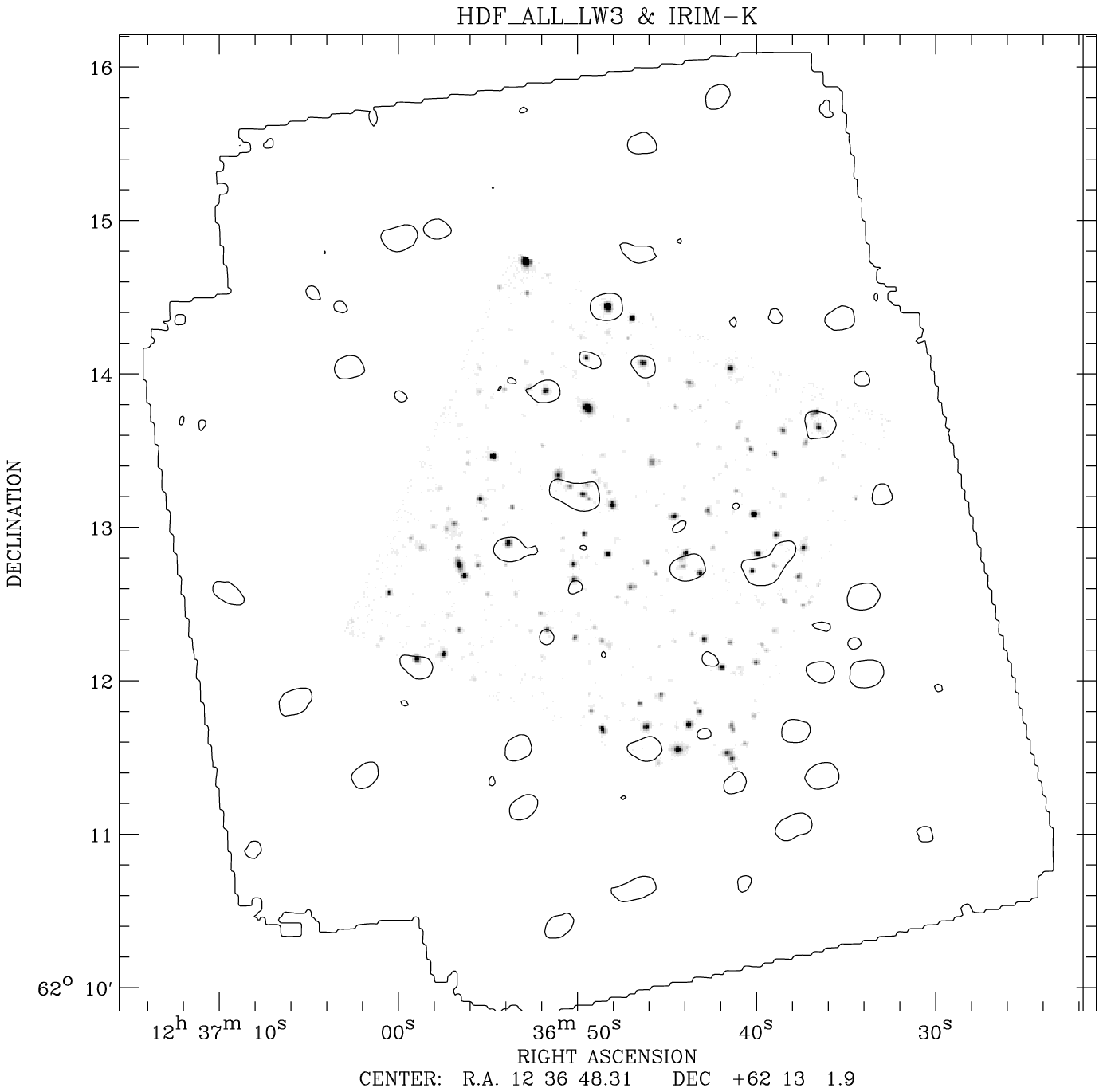}
  \caption{HDF K band near IR map (grey scale) overlaid with
the LW3 contour of the same 3 coadded rasters. The contour is at
about the 3~$\sigma$ level.
}
  \label{fi:klw3}
\end{figure*}

Essentially, the LW2 survey yields only two sources above the 4
$\sigma$ level of confidence.  They both have an LW3 couterpart.  We
conclude that at the 60 \microJy level, there are 2 (say at most a
few) sources in the 11 arcmin$^2$ main area of the HDF, a result at
odds with the numerous sources found by Goldschmidt \etal (1997). The
discrepancy has been addressed by Aussel \etal (1997). We note that
almost all LW2 sources have an LW3 counterpart. This small number
of sources is not in disagreement with the 15 sources found in a
different region but with a similar area by Taniguchi \etal (1997) at
a sensitivity level twice to 3 times better than here in the same LW2
filter.  The sensitivity level is here achieved with 1.7 hours of
integration per sky pixel of 3 arcseconds.  In the following, we
concentrate on the LW3 results which, because of the larger number of
sources, gives a better statistics on number counts. In the final
table (\ref{ta:hdfalllw3}), 34 objects are above the 4 $\sigma$ limit
of typically 100 to 150\microJy in an approximate area of 25
arcmin$^2$. The best sensitivity limit quoted is obtained with an
effective on-source integration time of 2.7 hours. Number counts are
discussed by Aussel \etal (1998).

The overall reliability of the method can now be assessed both
internally and externally:
\begin{itemize}
\item for the strong sources which are detected in individual rasters,
  the statistical significance in each of the subrasters (compare the
  signal-to-noise ratio S/N$_{1,2,3}$ to S/N$_t$ in
  Tables~\ref{ta:hdf2lw3}, \ref{ta:hdf3lw3}, \ref{ta:hdf4lw3}) is
  lower than the total result but the flux estimate is in agreement
  with the final map flux within the error bars (the quality factors are
  always large)
\item in the central common HDF area, the same sources are found in
  different \emph{completely independent} rasters. Indeed by comparing
  Tables~\ref{ta:hdf2lw3}, \ref{ta:hdf3lw3}, \ref{ta:hdf4lw3} and
  \ref{ta:hdfalllw3}, one can note the overall satisfactory
  photometric and astrometric consistency of the sources, within the
  stated error bars, for example HDF-2\_LW3\_9, HDF-3\_LW3\_5
  correspond to HDF\_ALL\_LW3\_25, and HDF-2\_LW3\_14, HDF-3\_LW3\_10,
  HDF-4\_LW3\_23 \\ agree with HDF\_ALL\_LW3\_33.  Strictly speaking,
  this argument does not hold for the sources that helped in
  correcting the astrometry.
\end{itemize}

The complete identification, as well as the detailed comparison with
the source list given by Goldschmidt \etal (1997), Rowan--Robinson
\etal (1997) and Aussel \etal (1998) and by radio surveys, as well as
the analysis of supplementary data taken as a repeat of LW2
observations will be dealt with in a forthcoming paper. A third external
level of consistency can already be made: more than a
third of the HDF LW3 sources are found in the VLA radio sample of
Richards et al (1998).  The analysis of Fig.~\ref{fi:vlw3} and
Fig.~\ref{fi:klw3} already reveals that, in the central HDF area, one
always finds one or several bright counterparts in the optical (B
magnitude less than 22).  An even more clear cut case is that the
sources can also be always associated with a K counterpart of
relatively bright 17 to 18 magnitude, except for one source
HDF\_ALL\_LW3\_20 at the lower west border with RA=12 36 46.2 and Dec=
62 11 33.5 (confirmed in LW3 observations by the 2 rasters HDF-3\_LW3
and HDF-4\_LW3).  This could be a good example of a galaxy very dim in
the optical (29th magnitude) and near-infrared domain relative to its
mid-infrared luminosity, for which no redshift can be measured
except in the mid, far infrared or radio domains. Another example is
HDF\_ALL\_LW3\_24 which is associated with a radio source identified
with a R$_{AB}$=23.9 elliptical galaxy by Richards \etal (1998).
Table~\ref{ta:hdfalllw3} gives the redshift (when available in the
internet lists, Cowie \etal 1998) of the galaxy which is the brightest
and nearest source (within 6 arcseconds) to a given ISOCAM source (in
a simple eye-ball sense). Almost all redshifts are within the 0.5 to 1
range. We are probably seeing the PAH spectral features around
$7.7\micron$ redshifted in the LW3 band. The fact that most of these
sources are not detected in LW2 (when observed) but detected in K
means that LW2 witnesses the break between stellar emission and
interstellar dust emission, for the ISOCAM HDF source redshifts.  Two
double sources can be spotted in this table. These are
HDF\_ALL\_LW3\_13 with 15, and 27 with 29.
Only one object is a star (at the border of the map) and not a galaxy:
HDF\_ALL\_LW3\_41 (see Aussel \etal 1998).

It is clear that the redundancy of the rasters plays a crucial role in
assessing the reliability of sources. The reliability of our method is
clearly demonstrated here on the HDF ISOCAM data, because of their
large (up to 50--100) redundancy. It will be used in various other
ISOCAM deep surveys, where such a redundancy cannot be afforded.  It
is shown here that the temporal triple beam-switch method plus a
classical spatial detection on an optimally coadded map of individual
measurements allow the source noise to be less than 2 parts per ten
thousand of the zodiacal background (as numerically found with the
present data). The method is linear and does not deal in different
ways with high and low flux sources (as long as they are small
compared with the zodiacal background).  This is achieved with a
modest-size telescope and modest-size detector array with a strong
reaction to cosmic rays and with some non-linear behaviours. Finally,
we note that the LW3 map is above the camera confusion limit by a
factor two.

\begin{acknowledgements}
  We wish to thank F. Boulanger, D. C\'esarsky, W. Reach and F.
  Vivares for their help during this project, and an anonymous referee
  for improving the manuscript.  Many discussions with L. Vigroux, D.
  Elbaz, H.  Aussel and J.--L. Starck along with ISOCAM Consortium
  members helped us with some issues.
\end{acknowledgements}

\end{document}